# The Critical Richardson Number and Limits of Applicability of Local Similarity Theory in the Stable Boundary Layer

Andrey A. Grachev • Edgar L Andreas • Christopher W. Fairall • Peter S. Guest • P. Ola G. Persson

---

Andrey A. Grachev (✉) • P. Ola G. Persson
NOAA Earth System Research Laboratory / Cooperative Institute for Research in Environmental Sciences, University of Colorado, 325 Broadway, R/PSD3, Boulder, CO 80305-3337, USA
e-mail: Andrey.Grachev@noaa.gov

Edgar L Andreas
NorthWest Research Associates, Inc., Lebanon, NH, USA

Christopher W. Fairall
NOAA Earth System Research Laboratory, Boulder, CO, USA

Peter S. Guest
Naval Postgraduate School, Monterey, CA, USA

**Abstract**

Measurements of atmospheric turbulence made over the Arctic pack ice during the Surface Heat Budget of the Arctic Ocean experiment (SHEBA) are used to determine the limits of applicability of Monin-Obukhov similarity theory (in the local scaling formulation) in the stable atmospheric boundary layer. Based on the spectral analysis of wind velocity and air temperature fluctuations, it is shown that, when both of the gradient Richardson number, $Ri$, and the flux Richardson number, $Rf$, exceed a 'critical value' of about $0.20 - 0.25$, the inertial subrange associated with the Richardson-Kolmogorov cascade dies out and vertical turbulent fluxes become small. Some small-scale turbulence survives even in this supercritical regime, but this is non-Kolmogorov turbulence, and it decays rapidly with further increasing stability. Similarity theory is based on the turbulent fluxes in the high-frequency part of the spectra that are associated with energy-containing/flux-carrying eddies. Spectral densities in this high-frequency band diminish as the Richardson-Kolmogorov energy cascade weakens; therefore, the applicability of local Monin-Obukhov similarity theory in stable conditions is limited by the inequalities $Ri < Ri_{cr}$ and $Rf < Rf_{cr}$. However, it is found that $Rf_{cr} = 0.20 - 0.25$ is a primary threshold for applicability. Applying this prerequisite shows that the data follow classical Monin-Obukhov local $z$-less predictions after the irrelevant cases (turbulence without the Richardson-Kolmogorov cascade) have been filtered out.

# 1 Introduction

Monin–Obukhov similarity theory (MOST) is the commonly accepted approach to describing near-surface turbulence in the stratified atmospheric boundary layer (ABL). According to MOST, any properly scaled statistics of the turbulence are universal functions of a stability parameter (Monin and Obukhov 1954)

$$\zeta \equiv z/L \tag{1}$$

defined as the ratio of a reference height $z$ and the Obukhov length scale (Obukhov 1946)

$$L = -\frac{u_*^3 \theta_v}{\kappa g <w'\theta_v'>}, \tag{2}$$

where $u_* = \sqrt{-<u'w'>}$ is the friction velocity, $\theta_v$ is the virtual potential temperature, $g$ is the acceleration due to gravity, $u$ and $w$ are the longitudinal and vertical velocity components, respectively, $[']$ denotes fluctuations about the mean value, and $<\ >$ is a time or space averaging operator (e.g., Sorbjan 1989; Garratt 1992; Wyngaard 2010).

Historically, the von Kármán constant $\kappa \approx 0.4$ is included in the definition of $L$ simply by convention. Specifically, the non-dimensional vertical gradients of mean wind speed, $U$, and potential temperature, $\theta$, according to the MOST are expressed as:

$$\varphi_m(\zeta) = \left(\frac{\kappa z}{u_*}\right)\frac{dU}{dz}, \tag{3a}$$

$$\varphi_h(\zeta) = \left(\frac{\kappa z}{T_*}\right)\frac{d\theta}{dz}, \tag{3b}$$

where $T_* = -<w'\theta'>/u_*$ is the temperature scale. The von Kármán constant $\kappa$ on the right-hand sides of Eqs. 3a, 3b is conventionally introduced solely as a matter of convenience such that $\varphi_m(0) = \varphi_h(0) = 1$ for neutral conditions ($\zeta \equiv 0$). Note that not all studies support the assumption $\varphi_h(0) = 1$ (e.g., Sorbjan 1989; Garratt 1992). The standard deviations of wind speed components $\sigma_\alpha$ and air temperature $\sigma_t$, in the frameworks of the MOST, are scaled as

$$\varphi_\alpha(\zeta) = \frac{\sigma_\alpha}{u_*}, \qquad (4a)$$

$$\varphi_t(\zeta) = \frac{\sigma_t}{|T_*|}, \qquad (4b)$$

where $\alpha$ (= $u$, $v$, and $w$) denotes the longitudinal, lateral, and vertical velocity components respectively. The exact forms of the universal functions (3) and (4) are not predicted by MOST and must be determined from field experiments.

It should be noted that originally MOST was based on the surface fluxes (surface scaling), where it is assumed that the turbulent fluxes are constant with height and equal to the surface values in the layer conventionally called a surface or constant-flux layer (roughly the lowest 10% of the entire ABL). According to Grachev et al. (2005), the constant-flux assumption underlying the original MOST is reasonably accurate for stable conditions in the range $0 < \zeta < 0.1$. Based on a second-order closure model, Nieuwstadt (1984) demonstrated that in the stable boundary layer (SBL) the assumption of height-independent fluxes is not necessary. He thus redefined Monin-Obukhov similarity in terms of local similarity (local scaling) for which the Obukhov length (Eq. 2) and the relationships (3) and (4) are based on the local fluxes at height $z$ (i.e., $z$-dependent fluxes) rather than on the surface values. Sorbjan (1986, 1988) argued that the functional forms of the universal functions (3) in the SBL are identical for both surface and local scaling assumptions. Thus, local scaling describes the turbulent structure of the entire SBL.

For future use, the gradient Richardson number, $Ri$, is defined by

$$Ri = \left(\frac{g}{\theta_v}\right)\frac{d\theta_v/dz}{(dU/dz)^2} = \frac{\zeta \varphi_h}{\varphi_m^2}, \tag{5}$$

while the flux Richardson number, *Rf*, (also known as the mixing efficiency) is

$$Rf = -\left(\frac{g}{\theta}\right)\frac{<w'\theta_v'>}{u_*^2(dU/dz)} = \frac{\zeta}{\varphi_m}, \tag{6}$$

where both *Ri* and *Rf* are expressed in a streamline coordinate system. The ratio of *Ri* to *Rf* is the turbulent Prandtl number defined by

$$Pr_t = \frac{K_m}{K_h} = \frac{<u'w'>(d\theta/dz)}{<w'\theta'>(dU/dz)} = \frac{Ri}{Rf} = \frac{\varphi_h}{\varphi_m}, \tag{7}$$

where $K_m$ and $K_h$ are the turbulent viscosity and the turbulent thermal diffusivity, respectively. The turbulent Prandtl number (7) describes the distinction in turbulent transfer between momentum (or downwind stress), $\tau = \rho u_*^2$, and sensible heat, $H_S = c_p \rho <w'\theta'>$. Turbulent momentum transfer is more efficient than turbulent heat transfer when $Pr_t > 1$ and vice versa.

Although MOST has provided relevant scaling tools for describing turbulence in the stratified ABL over the decades, precise limits of the applicability for similarity theory in stable conditions have not been established. In this paper, we use the extensive dataset from the Surface Heat Budget of the Arctic Ocean experiment (SHEBA) to determine the range of applicability of similarity theory (in the local scaling formulation) in the SBL.

**2 Local *z*-less Stratification**

In the very stable case ($\zeta \gg 1$), stratification inhibits vertical motions and the turbulence no longer communicates significantly with the surface (e.g., Monin and Yaglom 1971). In this case,

MOST predicts that $z$ ceases to be a scaling parameter; that is, various quantities become independent of $z$ (Obukhov 1946; Monin and Obukhov 1954). This limit requires that $z$ cancels in Eqs. 3a, 3b, which leads to linear relationships for the stability functions:

$$\varphi_m(\zeta) = \beta_m \zeta ,  \qquad (8a)$$

$$\varphi_h(\zeta) = \beta_h \zeta ,  \qquad (8b)$$

where $\beta_m$ and $\beta_h$ are numerical coefficients. Earlier Wyngaard and Coté (1972) and Wyngaard (1973) termed the limit (8) "local $z$-less stratification" (height-independent); dimensionless standard deviations (4) become asymptotically constant in $z$-less stratification. Dias et al. (1995) found that properly normalized third-order moments remain constant with height and are hence in good agreement with the $z$-less concept.

A simple linear interpolation provides blending between neutral and $z$-less cases (e.g. Zilitinkevich and Chalikov 1968; Webb 1970)

$$\varphi_m(\zeta) = 1 + \beta_m \zeta ,  \qquad (9a)$$

$$\varphi_h(\zeta) = 1 + \beta_h \zeta ,  \qquad (9b)$$

which implies $Pr_t(0) = \varphi_h(0)/\varphi_m(0) = 1$. Relationships (9) would be also the linear approximations if each function (3) is expanded into power series for fairly small values of $\zeta$ (Garratt 1992) and reduce to (8) in the asymptotic limit $\zeta \to \infty$. Traditionally the linear equations in the stable case (9) together with the relations for $\varphi_m(\zeta)$ and $\varphi_h(\zeta)$ in the unstable case are called the Businger–Dyer relations (Dyer and Hicks 1970; Businger et al. 1971; Dyer 1974; Businger 1988). Since the landmark 1968 Kansas field experiment (Businger et al. 1971), the linear equations (9) have fit the available experimental data well for $\zeta < 1$ and measurements suggest $\beta_m \approx \beta_h \approx 5$ (Högström 1988; Sorbjan 1989; Garratt 1992; Handorf et al. 1999; Wyngaard 2010).

Although in many cases, MOST works reasonably well, z-less scaling (8) was questioned for $\zeta > 1$ based on analyses of current extensive datasets. Forrer and Rotach (1997), Howell and Sun (1999), Yagüe et al. (2001, 2006), Klipp and Mahrt (2004), Cheng and Brutsaert (2005), Baas et al. (2006), and Grachev et al. (2005, 2007a, 2008) reported that the stability functions (3) increase more slowly with increasing stability than is predicted by the linear equations (9) and, moreover, one ($\varphi_h(\zeta)$) or both functions become approximately constant in very stable conditions. Based on an analysis of standard deviations (4), Pahlow et al. (2001) found that relationships (4) do not follow the z-less predictions (see also the survey in Hong 2010). Thus, the situation is quite serious because several independent groups came to the same conclusion based on data collected in a variety of conditions.

Several studies recently attempted to remove this ambiguity associated with the z-less concept. Basu et al. (2006) revisited the data used by Pahlow et al. (2001) and applied a wavelet-based filter to this and several other datasets to remove non-turbulent effects. Their analysis supported the validity of z-less stratification for $\sigma_u / u_*$, and they concluded that the Pahlow et al. (2001) results regarding the invalidity of z-less stratification under very stable conditions were biased by the inclusion of non-turbulent motions. Hong et al. (2010) nevertheless supported the validity of z-less stratification up to only $\zeta \approx 0.5$ by applying the Hilbert-Huang transform to separate turbulent from non-turbulent motions.

Mahrt (2007) analyzed extensive eddy-correlation datasets to examine the influence of non-stationarity of the mean flow on the flux-gradient relationships for weak and moderate stability, $\zeta < 1$. He found that non-stationarity causes a 'leveling-off' of $\varphi_m$ and to a lesser extent of $\varphi_h$. However, even for stationary cases, bin-averaged $\varphi_m$ data support the first linear equation (8) with $\beta_m = 5$ up to only $\zeta \approx 0.6$; while in the range $0.6 < \zeta < 1$, the function $\varphi_m$ increases more slower than the linear prediction (Mahrt 2007, Fig. 3).

Kouznetsov and Zilitinkevich (2010) pointed out that 'levelling-off' of $\varphi_m$ is also associated with the fact that experimental data contain points for which the flux Richardson number (6) is > 1. In these cases, buoyancy consumes more energy than wind shear produces, a combination that leads to non-stationarity (decaying turbulence).

These contradictory results indicate that the validity of z-less stratification is still an open question that requires further clarification.

**3 The SHEBA Dataset**

Turbulent measurements made over the Arctic pack ice during the Surface Heat Budget of the Arctic Ocean experiment (SHEBA) took place in the Beaufort Gyre from October 1997 through September 1998. Andreas et al. (2006; 2010a, b), Persson et al. (2002) and Grachev et al. (2005, 2007a, 2007b, 2008) describe the SHEBA site and review flux and profile measurements, data processing, accuracy of measurements, and instrument calibration. Here we provide relevant information about the turbulent measurements.

Turbulent fluxes, variances, and mean meteorological data were continuously measured on a 20-m main tower at five levels, hereafter level 1 (nominally $z_1 \approx 2.2$ m above the surface), level 2 ($z_2 \approx 3.2$ m), level 3 ($z_3 \approx 5.1$ m), level 4 ($z_4 \approx 8.9$ m), and level 5 ($z_5 \approx 18.2$ or 14 m during most of the winter). Thus, the reference height $z$ in (1) and (2) is $z \equiv z_n$ ($n = 1$–5). Each level of the tower was instrumented with identical Applied Technologies, Inc. (ATI), three-axis sonic anemometer/thermometers (K-probe) that sampled at 10 Hz and a HMP-235 temperature and relative humidity (T/RH) probe. An Ophir fast-response infrared hygrometer was mounted on a 3-m boom at an intermediate level (about 8 m) just below level 4. Although a sonic anemometer measures the so-called 'sonic' temperature, which is close to the virtual temperature, the moisture correction in sonic temperature is usually small for Arctic conditions (Grachev et al. 2005, p. 205).

The 'slow' T/RH probes provided air temperature and relative humidity measurements at five levels and were used to evaluate the vertical temperature and humidity gradients. The mean wind speed and wind direction were derived from the sonic anemometers, with rotation of the anemometer axes needed to place the measured wind components in a streamline coordinate system. We used the most common method, which is a double rotation of the anemometer coordinate system, to compute the longitudinal, lateral, and vertical velocity components in real time (Kaimal and Finnigan 1994, Sect. 6.6). The 10-Hz raw data were first edited to remove spikes from the data stream. Turbulent fluxes and appropriate variances at each level are based on 1-hr averaged data, and to obtain 1-hr averaged values, cospectra and spectra were normally computed from seven overlapping 13.65-min data blocks (corresponding to $2^{13}$ data points) and then averaged within an hour (see other details in Persson et al. 2002).

Turbulent covariance and variance values were derived through the frequency integration of the appropriate hourly-averaged cospectra and spectra. To separate the contributions from mesoscale motions to the calculated eddy-correlation flux, a low-frequency cut-off at 0.0061 Hz (the sixth spectral value or a period of about 3 min) was applied on the cospectra as a lower limit of integration; the upper limit of integration is 5 Hz (the Nyquist frequency). The low-frequency cut-off for turbulent contributions is chosen to lie in the spectral gap between the small- and large-scale contributions to the total transport (see spectra and cospectra plots in Grachev et al. (2005, Fig.8) and Grachev et al. (2008, Fig. 3)). Our cut-off value is also consistent with the low-frequency cut-off at 0.005 Hz chosen by Kaimal et al. (1972) and 0.008 Hz used by Caughey (1977) for the case of the SBL.

Several data-quality indicators based on objective and subjective methods have been applied to the original flux data (e.g., Grachev et al. 2007a, p. 319). In particular, data with a temperature difference between the air (at median level) and the snow surface less than 0.5°C have been omitted to avoid the large uncertainty in determining the sensible heat flux. To avoid a possible flux loss caused by inadequate frequency response and sensor separations, a prerequisite that $U > 1$ m s$^{-1}$ has also been imposed.

Vertical gradients of the mean wind speed, potential temperature, and specific humidity that appear in $\varphi_m$, $\varphi_h$, $Ri$, and $Rf$ were obtained by fitting a second-order polynomial through the 1-hr profiles followed by evaluating the derivative with respect to $z$ for levels 1–5 (Grachev et al. 2005, their Eqs. 8). However Grachev et al. (2007b) show that the results are not sensitive to the method of evaluating the wind speed and temperature gradients (cf. their Figs. 1 and 2).

The SHEBA site was located on Arctic pack ice, which had no large-scale slopes or heterogeneities; the site was a few hundred kilometers from land and thus provided almost unlimited and extremely uniform fetch. Only surface roughness features of less 1 km length scale and less than about 3 m in height (e.g., leads, ice ridges, ice rubble, and smooth snow-covered ice) are present for many hundreds of kilometers. For these reasons, the SHEBA flux data are not generally contaminated by drainage (katabatic) flows, strong local advective flows, or orographically-generated gravity waves. Thus the SBL observed most often during SHEBA can be characterized as a traditional stable boundary layer (e.g., Banta et al. 2006, their Fig. 1). According to Mahrt and Vickers (2002), the traditional or surface-flux dominated SBL is defined as a stably stratified layer where turbulence is generated by surface roughness, and momentum and heat fluxes are approximately constant or decrease monotonically with height.

**4 Turbulence Decay in the SBL**

4.1 Analysis of Turbulence Spectra and Cospectra

Figure 1 shows typical one-dimensional, raw energy spectra of the longitudinal, lateral, and vertical velocity components, and the sonic temperature at five levels for weakly and moderately stable conditions on 5 December 1997 (YD 339.375 UTC). Appropriate raw cospectra of the downward momentum flux (or longitudinal wind-stress component) and the sonic temperature flux are shown in Fig. 2. Both spectra and cospectra for the weakly and moderately stable conditions are more or less regular and not contaminated by mesoscale motions.

According to Fig. 1, the turbulent spectral curves have a wide credible inertial subrange, which displays the $-5/3$ Kolmogorov power law at high frequencies (corresponding to the slope of $-2/3$ for the frequency-weighted spectra plotted in Fig. 1). Although stability increases with increasing height, this regime is associated with Kolmogorov turbulence at all five sonic levels. In the low frequency subrange, the spectra exhibit a power law with an exponents close to $+1$ (cf. Caughey 1977; Kaimal 1973; Kaimal and Finnigan 1994). Note that the spectral energy at low frequencies for the vertical velocity component (Fig. 1c) is much less than for the horizontal components (Figs. 1a, b).

Frequency-weighted cospectra in Fig. 2 are plotted in log-linear coordinates, so that the area under the spectral curve represents the total covariance. According to Fig. 2, the magnitudes of the stress and the sensible heat flux at levels 1 and 2 are approximately the same, and then flux magnitudes monotonically decrease with increasing height. Thus, levels 1 and 2 are within the surface or constant-flux layer, where the original MOST is applicable (surface scaling), whereas Nieuwstadt's (1984) local scaling should be applied for levels 3-5. Values of the turbulent fluxes and the stability parameters (1), (5), and (6) are listed in the figure captions. Additional plots of the spectra and cospectra for weakly and moderately stable conditions can be found in Grachev et al. (2008, Fig. 2) and Grachev et al. (2005, Fig. 3), respectively. With increasing stability, the vertical fluxes and the turbulence intensity decrease because negative buoyancy inhibits vertical transfer. Spectra and cospectra for more strongly stable conditions are shown in Figs. 3 and 4 (data obtained on 30 December 1997, YD 364.0833 UTC). Additional plots for this regime can be found in Grachev et al. (2005, Fig. 8) and Grachev et al. (2008, Fig. 3).

According to Fig. 3, the surface-generated wind shear is still large enough to maintain small-scale turbulence at the three near-surface levels and the inertial subrange is identifiable in spectra at these levels. Small-scale turbulence has obviously collapsed at level 5 in both wind velocity and temperature spectra, and the inertial subrange associated with a Richardson-Kolmogorov cascade disappears (Fig. 3). One may assume that the regime observed at level 4 (Figs. 3 and 4) represents a transitional case from Kolmogorov turbulence at levels 1-3 to the regime that occurs at the level 5. Furthermore, often spectral densities in the high-frequency range for this regime increase with increasing frequency and show a slope close to unity that is associated with 'white noise' in the sensor (see Fig. 3 in Grachev et al. 2008). In contrast to the weakly and moderately stable conditions, the low-frequency disturbances in the spectra and cospectra for very stable conditions are more pronounced (except the *w*-component); however, according to the SHEBA data, the low-frequency part is usually separated by a spectral gap from the high-frequency turbulence in this regime for measurements at the lower levels (see Grachev et al. 2005, Fig. 8; Grachev et al. 2008, Fig. 3). Similar behaviour of spectra and cospectra for different stabilities has been observed in laboratory experiments according to Rohr et al. (1988).

Thus, we observed a layered structure with weak turbulence occupying the near-surface layer (usually the two to three lowest sonic levels) and collapsed turbulence (no small-scale turbulence) above (the upper one or two sonic levels). A similar very shallow SBL, where a near-surface layer with continuous turbulence may be less than 5 m deep, has been reported by Smedman (1988), King (1990), Mahrt and Vickers (2006) and Banta et al. (2007).

4.2 The Critical Richardson Number

One of the fundamental problems of stably stratified shear flows is associated with the existence of a threshold beyond which the turbulence is suppressed. Up to now, the transition from turbulent to laminar flow and vice versa remains an enigma (see Galperin et al. 2007, for discussion). Since Richardson (1920), it has been assumed that, when the gradient Richardson number (5) exceeds some critical value, $Ri_{cr}$, turbulence collapses to laminar flow. Richardson (1920) predicted that for $Ri > Ri_{cr} = 1$ no turbulence would survive. According to Miles (1961) and Howard (1961), laminar, steady, inviscid flow with $Ri > 0.25$ everywhere in the flow will remain stable to small perturbations. This is a sufficient condition for the stability of a sheared, stratified flow that was later verified experimentally by Scotti and Corcos (1972), who found $Ri_{cr} = 0.22$.

However, these result do not necessarily imply that existing turbulence disappears as $Ri$ increases above $Ri_{cr} = 0.25$. Laboratory observations of turbulence by Rohr et al. (1988) in a salt-stratified water channel show that turbulence survives for $Ri > Ri_{cr} \approx 0.25$ but turbulence growth is suppressed. Itsweire et al. (1993) compared a direct numerical simulation for stratified turbulence to the laboratory experiments of Rohr et al. (1988). The numerical results show that, if $Ri > Ri_{cr} \approx 0.21$ (which corresponds to $Rf_{cr} \approx 0.2$-$0.25$), turbulence decays in good agreement with the laboratory measurements of Rohr et al. (1988). Van de Wiel et al. (2007) analytically and numerically studied the collapse of turbulence in a stably stratified plane channel flow and argued that $Ri_{cr} = 0.2$. Laboratory measurements in the wind tunnel by Ohya et al. (2008) show that $Ri_{cr} \approx 0.25$ describes the transition from turbulence bursting events to quiescent periods and vice versa (see their Fig. 16). Other published estimates of $Ri_{cr}$ are summarized in Zilitinkevich and Baklanov (2002, their Table I).

Note that the problem of the turbulent-laminar transition (or from turbulent to non-turbulent flow in the general case) presently has no rigorous mathematical treatment. Simplified analyses are based primarily on the equations of turbulent kinetic energy (TKE) and mean square temperature fluctuations and postulate that a critical condition for the transition from turbulence to laminar flow is associated with the flux Richardson number (6). Ellison (1957) first used this approach and presented an analysis that suggests the threshold $Rf_{cr} = 0.15$. Further, Townsend (1958) found that continuous turbulence cannot be maintained if $Rf > 0.5$; that is, he found it impossible to obtain solutions that satisfy the equations for TKE and temperature fluctuations beyond $Rf_{cr} = 0.5$. Arya (1972) critically reconsidered the theoretical models of Ellison (1957) and Townsend (1958) and found $Rf_{cr} = 0.15$-$0.25$ using field data available at the time.

Estimates of the critical flux Richardson number based on turbulence closure models yield $Rf_{cr}$ = 0.21 (Mellor 1973, see Eq. 43a for $\varphi_m$) and $Rf_{cr}$ in the range 0.18-0.27 for different empirical constants (Yamada 1975). Yamada (1975) also pointed out a simple derivation of $Rf_{cr}$ based on the definition (6) and Eq. 8a: that is, $Rf \to Rf_{cr} = 1/\beta_m \approx 0.2$ for $\beta_m \approx 5$ as $\zeta \to \infty$ (cf. Garratt 1992, Sect. 3.3.1; Wyngaard 2010, Eq. 12.38). Recently Zilitinkevich et al. (2010) also suggested $Rf_{cr} = 0.2$ based on the TKE budget equation but, at the same time, they argued that there is no critical value for the gradient Richardson number.

The critical Richardson number derived from measurements in the atmospheric SBL is a controversial issue (see the survey by Andreas (2002) and references therein). Although all atmospheric observations show that, with increasing stability, turbulent fluctuations decrease, become more intermittent, and finally disappear, different thresholds beyond which the turbulence disappears have been reported. For example, Okamoto and Webb (1970) and Busch (1973) found that $Ri_{cr} \approx 0.2$ based primarily on the analysis of turbulence spectra. Observations in the atmospheric SBL by Kondo et al. (1978) show that flow is fully turbulent and relationships (9) are valid for $Ri < 0.24$ although intermittent turbulence can persist beyond this value up to $Ri = 1$. According to Kondo et al. (1978), beyond $Ri = 2$ turbulence almost ceases. Mauritsen and Svensson (2007) analyzed tower-based turbulence observations from six different datasets (including an earlier version of the SHEBA dataset) and found that weak turbulent mixing persists beyond the critical gradient Richardson number (their Figs. 2 and 4). However, the velocity field becomes highly anisotropic and is dominated by horizontal motions.

It is worth noting that an alternative approach for evaluating the critical Richardson number from atmospheric observations is associated with a formulation for the height of the SBL (Handorf et al. 1999; Zilitinkevich and Baklanov 2002; Vickers and Mahrt 2004). The apparent range in critical Richardson numbers cited above can be explained in part by hysteresis (e.g., Stull 1988, Sect. 5.6.2). Turbulent flow at increasing $Ri$ becomes non-turbulent (termination of turbulence) when $Ri$ becomes larger than around 1; while at decreasing $Ri$, non-turbulent flow becomes turbulent (onset of turbulence) when $Ri$ drops below $Ri_{cr} \approx 0.20 - 0.25$ (Woods 1969; Nilsson 1996).

As the Richardson number approaches its canonical 'critical value' of 0.20 or 0.25, turbulence decays and vertical fluxes become small. Figures 5 and 6 show decaying momentum flux (downwind stress) and sensible heat flux as a function of the different stability parameters. Unlike the momentum flux, which decreases monotonically, the downward heat flux plotted versus stability parameters has a minimum (e.g., Derbyshire 1990; Jordan et al. 1999; Grachev et al. 2005). Recall that the fluxes in Figs. 5 and 6 are computed over the high frequency part of the turbulence cospectra with about a 3-min cut-off time scale as the low-pass filter (see Sect. 3).

According to the SHEBA data, both the small-scale wind stress and the sensible heat flux decrease rapidly with increasing stability, but small-scale turbulence exists even when both $Ri$ and $Rf$ exceed 0.20 – 0.25. However, according to Figs. 5 and 6, the turbulent wind stress falls off faster with increasing stability than the sensible heat flux (see also Grachev et al. 2008 their Fig. 6). The small-scale momentum flux persists up to $Ri \approx 0.3$, $Rf \approx 0.4$, and $\zeta \approx 10$; whereas the small-scale sensible heat flux persists up to at least $Ri \approx 2$, $Rf \approx 3$, and $\zeta \approx 50$. Note that the sensible heat flux data have a larger scatter in the supercritical regime that complicates estimating a threshold for $H_S$. Beyond these limits, the turbulent fluxes are negligible, and non-zero high-frequency values are most probably associated with sensor noise (see Figs. 3 and 4 for level 5). In very stable conditions, the physical fluctuations of wind speed and sonic temperature are small and approach the resolution of the ATI sonic anemometer (0.01 m s$^{-1}$ and 0.01°C). This resolution leads to a step ladder appearance in the data time series, and the turbulent fluxes cannot be reliably calculated (Vickers and Mahrt, 1997, their Fig. 1b).

Thus, our observations show that there is no clear 'critical' Richardson number for decaying turbulent fluxes (Figs. 5 and 6). This result is consistent with the conclusion of Galperin et al. (2007) that a single critical Richardson number for the suppression of turbulence does not exist. We may even speculate that $Ri$ does not reach a critical value at all and turbulence (presumably in the low-frequency range) does not cease as stability increases. Note that the data presented in Figs. 5 and 6 combine both cases when stability is increasing from smaller to larger values (decaying turbulence) and cases when stability is decreasing from larger to smaller values (growing turbulence). One may speculate that the separation of these two cases gives different behaviour of the turbulent fluxes in the vicinity of the critical Richardson number (hysteresis effect).

Zilitinkevich et al. (2007) were the first to show that a second-order moment closure model, which they called an "energy and flux budget turbulence closure model", can predict the persistence of turbulence beyond the critical Richardson number; thus, no critical Richardson number exists. Because the model of Zilitinkevich et al. (2007) differs significantly from the 'mainstream' of second-order moment closure models, Canuto et al. (2008) modified a traditional second-order moment closure model and confirmed the findings of Zilitinkevich et al. (2007); that is, simulated turbulence can exist at any $Ri$. Furthermore, the recent third-order moment model of Ferrero et al. (2011) also predicts the persistence of turbulence beyond the critical Richardson number. In contrast, a simple two-equation turbulence closure of Baumert and Peters (2004) and Peters and Baumert (2007) predicts the fully stationary state of turbulence at $Ri = Rf = 0.25$ with turbulence collapsing at $Ri = 0.5$.

Turbulence in the supercritical regime may have a different nature. One may suggest that fluxes in the supercritical regime shown in Figs. 5 and 6 represent surviving small-scale turbulence; that is, this is turbulence that passed through the critical point as stability increased and did not dissipate completely. It seems that the small-scale turbulence at level 4 in Figs. 3 and 4 is an example of such a case whereas the regime observed at level 5 is a case of collapsed small-scale turbulence. The surviving turbulence may be associated with mechanical or thermal 'inertia' ('turbulent inertia') during the passage through the critical point, and therefore it may well be non-stationary turbulence. Significant turbulence events in the supercritical regime can also be generated or modulated by gravity waves, wave-wave interactions, dynamic or thermodynamic instability, microfronts, and numerous other complex structures (see Mahrt 2010a, 2010b, 2011, and references therein).

4.3 Upper Limit of Applicability of the –5/3 Kolmogorov Power Law in the SBL

According to Figs. 1 and 3 (see also Figs. 2 and 3 in Grachev et al. 2008), both gradient and flux Richardson numbers generally increase with increasing height and, at some level, reach a critical value when the –5/3 Kolmogorov power law fails. In Figs. 7 and 8, we plot spectral slopes of velocity components and sonic temperature in the inertial subrange to clarify this issue. These figures show variations of the spectral slope for the longitudinal, lateral, and vertical velocity components and for the sonic temperature in the inertial subrange plotted versus $Ri$ (Fig. 7) and $Rf$ (Fig. 8). Spectral slopes were computed in the frequency domain 0.96 – 2.95 Hz (cf. Figs. 1 and 3). To obtain the slope within this range, six individual overlapping slopes have been computed between the 44th and 50th, 45th and 51st, …, and 49th and 55th spectral values; and then the median of these six values was taken as the representative slope (the total frequency domain in spectral or cospectral plots contains 60 points). With this procedure, we avoided the influence of possible spectral spikes on the estimation of the inertial-range spectral slope (see Figs. 1 and 3).

As expected, the $-5/3$ Kolmogorov power law for the inertial subrange in the spectral density of velocity fluctuations is obeyed over a wide range of the stability parameters. However, according to Figs. 7 and 8, as both $Ri$ and $Rf$ reach a value of $0.20 - 0.25$ (which corresponds to $\zeta \approx 1$), the spectral slope abruptly deviates from the classical value of $-5/3$ and increases and becomes more variable with increasing stability. Eventually, in the very stable regime, the spectral slope levels off asymptotically at zero which corresponds to 'white noise' in the sensor (see spectra at level 5 in Fig. 3, Grachev et al. 2008). Note that, according to Figs. 7 and 8, a break-up of the $-5/3$ Kolmogorov power law for the $w$ component occurs a little earlier than for the horizontal components, evidence of a violation of the local isotropy hypothesis.

The spectral slope of the sonic temperature fluctuations in the inertial subrange behaves quite differently (Figs. 7d and 8d). In all these bin-averaged plots, the temperature spectrum has only a modest $-5/3$ regime. This regime is the scalar analogue of Kolmogorov's $-5/3$ law for velocity components and was predicted by Obukhov and by Corrsin, who independently extended Kolmogorov's arguments to passive scalars (e.g. Garratt 1992; Wyngaard 2010). According to the bin-averaged data in Figs. 7d and 8d, a credible $-5/3$ power-law region is generally associated with the spectra measured at near-ground levels. For upper levels, on the other hand, the averaged spectral roll-off rates are less steep than the expected $-5/3$.

The scalar spectrum in geophysical shear flows for neutral stratification is ambiguous because of small temperature gradients; as a result, the spectral slope asymptotically tends to zero as the stability parameters $Ri$, $Rf$, and $\zeta$ approach zero. However, the data presented in Figs. 7d and 8d do not contradict the conclusion that the $-5/3$ power law in the temperature spectrum also fails when $Ri$ and $Rf$ exceed a critical value around $0.20 - 0.25$. Note also that a spectral slope plotted versus $\zeta$ increases more smoothly (not shown) than in the plots using $Ri$ (Fig. 7) and $Rf$ (Fig. 8), and one may suggest that there is no 'critical' value of $\zeta$ for which the inertial subrange disappears. Notice, currently we are unable to say with certainty that a single "critical point" located in the narrow range $0.20 - 0.25$ exists or whether there is a gradual transition within this range.

Thus, according to Figs. 7 and 8, the upper limit of applicability of the $-5/3$ Kolmogorov power law (the Obukhov-Corrsin law for the passive scalar) and therefore for the existence of the inertial subrange in stably stratified layers is described by the inequalities

$$Ri < Ri_{cr}, \qquad (10a)$$

$$Rf < Rf_{cr}, \qquad (10b)$$

where both critical values $Ri_{cr}$ and $Rf_{cr}$ are about 0.20-0.25. It is notable that the critical values of $Ri$ and $Rf$ that describe a break-up of Kolmogorov turbulence (Figs. 7 and 8) coincide with their canonical values 0.20 or 0.25 but are derived from quite different reasoning.

The data presented in Figs. 7 and 8 do not allow us identify a threshold more accurately than $0.20 - 0.25$. The case study shown in Fig. 3 suggests that the critical value may be closer to 0.25 than to 0.20 (see figure caption); however, both $Ri_{cr}$ and $Rf_{cr}$ in (10) are averaged values in principle because of the non-deterministic nature of turbulence. Some high values of $Ri$ and $Rf$ (including 'supercritical' values) at level 1 and, in part, at level 2 for the cases shown in Figs. 1 and 3 (see figure captions) are outliers because levels 1 and 2 are located too close to the surface (i.e., within the roughness or blending sublayer). Here, consequently, the vertical gradients are more affected by surface heterogeneity (see discussion in Grachev et al. 2005 on p. 217 and their Fig. 16). In these cases, *Ri* and *Rf* are misleading as stability parameters.

It should be noted that, for field data analysis, for example, to estimate the dissipation rate of TKE derived from the –5/3 subrange in velocity spectra, both inequalities (10) should be used to filter out all irrelevant cases. This requirement stems from the fact that random data scatter is inherent in atmospheric turbulence measurements and using only one criterion, for example Eq. 10a, would be incomplete. Figure 9 shows a plot of the bin-averaged gradient Richardson number versus the flux Richardson number to clarify this point. According to Fig. 9, using only one inequality, (10a) or (10b), may skip some undesired data points located in the second and the fourth quadrants denoted by the dashed lines. Figure 9 is plotted as *Ri* against *Rf* rather than vice versa to avoid an outlier problem in plots where *Ri* is the independent variable, as briefly described in Grachev et al. (2008, p. 159-160) and discussed in detail by Grachev et al. (2012). To limit the influence of outliers on bin-averaging in the plots where *Ri* is used as the independent variable (e.g., in Figs. 5-7), a prerequisite has been imposed on the data in the form $0.5 < Ri / Ri_{SHEBA} < 2$ (Grachev et al. 2008, 2012), where $Ri_{SHEBA} = \zeta \varphi_{h\ SHEBA} / \varphi_{m\ SHEBA}^2$ is based on the SHEBA profile functions specified in Grachev et al. (2007a).

Note that a plot of *Ri* versus *Rf* can be used also for evaluating the turbulent Prandtl number because $Pr_t$ is the slope of the line connecting a data point and the origin in such a representation (see Eq. 7). This formulation contains only one shared variable, *dU/dz*, and therefore is less affected by self-correlation compared to relationships where $Pr_t$ is plotted versus $Ri$ or $Rf$ (see discussion in Grachev et al. 2007b). According to Fig. 9, a majority of points is located below the diagonal line, that is, in the region $Ri < Rf$ (or $Pr_t < 1$).

Unfortunately, there are not many observations of the spectral behaviour in very stable conditions. Early measurements by Okamoto and Webb (1970) show that, for $Ri > 0.2$, several runs produced near-zero slope whilst others gave spectral slopes between –1.2 and –2.5. That is, the spectrum changed unpredictably from one occasion to another. Caughey (1977) found that, in general, for $Ri$ between 0.2 and 0.3, the turbulence was weak and the spectral estimates fell to near noise levels. Busch and Panofsky (1968) reported that, in regions over which the spectra obey the –5/3 power law, the ratio of the lateral to the longitudinal spectra shows fair agreement with the 4/3 ratio predicted by the Kolmogorov hypothesis for the inertial subrange (the vertical-longitudinal ratio has a similar tendency). Based on the data collected during the 1968 Kansas experiment, Kaimal (1973) presents evidence for isotropy in the inertial subrange for $0.05 < Ri < 0.2$: namely, the –5/3 power law for velocity spectra and the 4/3 ratio of the spectral energies in the transverse and longitudinal components. However, at very high stability (Kaimal 1973, Table 1, run 23), a reported ratio between vertical and longitudinal velocity components that is less than 4/3 shows signs of anisotropy.

The conclusion that the Richardson-Kolmogorov cascade mechanism is unsuitable in the supercritical regime has practical significance for turbulence closure models and large-eddy simulation, which uses, explicitly or implicitly, Kolmogorov's relation for obtaining the dissipation rate of turbulent kinetic energy (e.g. Jimenez and Cuxart 2005).

**5 Similarity Functions in Subcritical and Supercritical Regimes**

Observations in the ABL have shown that properly normalized spectra and cospectra can be represented by a series of universal curves based on Monin-Obukhov scaling such that the inertial subrange collapses to a single straight line with a −2/3 slope on a log-log scale (Kaimal et al. 1972; Kaimal and Finnigan 1994). In the case of the SBL, different frequency-weighted spectra and cospectra collapse into single universal curves not only in the inertial subrange but over much of the frequency range when they are normalized by their variances or covariances, respectively, and plotted against the non-dimensional frequency $f/f_o$. Here $f_o$ is the frequency at which the extrapolated −2/3 inertial subrange slope of the non-dimensional spectrum or cospectrum intersects a horizontal line at one (Kaimal 1973; Caughey 1977; Kaimal and Finnigan 1994, Fig. 2.10). The frequency scale $f_o$ is closely associated with the frequency of the spectral or cospectral maximum (Kaimal and Finnigan 1994).

Obviously, the Kolmogorov inertial subrange is inseparably linked to the spectral maximum associated with the energy-containing turbulent eddies. In Kaimal's (1973) model, this is realized through the frequency scale $f_o$. Existence of the energy cascade requires permanent energy pumping at the energy-containing scales. A collapse of the energy-containing eddies causes the inertial subrange to collapse and leads to violations of spectral and cospectral similarity. The turbulent fluxes and variances are derived through frequency integration of the appropriate cospectra and spectra, mainly in the frequency band associated with the energy-containing turbulent eddies. Thus, violation of spectral and cospectral similarity leads to violation of flux-profile similarity. In other words, the applicability limit (10) of the −5/3 Kolmogorov power law should also be the applicability limit of local similarity theory.

We revisited traditional plots of the scaled quantities (3) and (4) in the light of the above applicability condition (10). Plots of $\varphi_m$ and $\varphi_h$ versus $\zeta$ for all the stable SHEBA data are shown in Figs. 10a and 11a, respectively. According to Figs. 10a and 11a, both stability functions (3) increase more slowly with increasing $\zeta$ than is predicted by the linear equations (9), a result that brings into question $z$-less scaling (8). As mentioned in Sect. 2, similar results have been obtained in other studies. We assume that a reason for this behaviour in $\varphi_m$ and $\varphi_h$ is our including in the analysis data points for which Monin-Obukhov theory may not be applicable. While this view is well known (e.g., Basu et al. 2006; Hong et al. 2010), determining a threshold beyond which Monin-Obukhov similarity fails is a non-trivial problem.

We have tested various constraints imposed on the data based on the inequalities (10). Figures 10b and 11b show the same plots as in Figs. 10a and 11a but when both prerequisites (10) with $Ri_{cr} = Rf_{cr} = 0.2$ have been imposed on the individual data for all five levels and medians (yellow symbols). According to Figs. 10b and 11b, the situation has been dramatically improved; that is, the SHEBA data almost perfectly follow Eq. 9 with $\beta_m = \beta_h = 5$ (especially for $\varphi_m$); and, therefore, the data are consistent with the $z$-less predictions (8) in the limit $\zeta \to \infty$. The data for $\varphi_h$ at level 1 and partially at level 2 (Fig. 11b) show systematic bias because these levels are located too close to the surface and the data are more affected by surface heterogeneity, as mentioned in Sect. 4.

The above approach is critical not only for identifying cases where similarity theory may be valid (Figs. 10b and 11b), but also it is important for studying turbulence in the supercritical regime ('supercritical' turbulence). Figures 10c and 11c show behaviour of the non-dimensional gradients (3) when a prerequisite $Ri_{cr} > 0.2$ and $Rf_{cr} > 0.2$ has been imposed on the data. This prerequisite selects small-scale, non-Kolmogorov turbulence (or more precisely 'turbulence without Richardson-Kolmogorov cascade', cf. Mazellier and Vassilicos 2010) that survived the transition through the critical point.

According to Figs. 10c and 11c, both $\varphi_m$ and $\varphi_h$ increase linearly in the range approximately $0.1 < \zeta < 3$. One may assume that the $z$-less concept (8) is also applied to the turbulence in the supercritical regime, but the numerical coefficients $\beta_m$ and $\beta_h$ should be different than in the subcritical case (Figs. 10b and 11b). Deviation from a linear dependence for $\zeta > 3$ in Figs. 10c and 11c may be eliminated by imposing the prerequisite $0.2 < Rf < 0.5$ on the data. Applying this prerequisite extends the linear ($z$-less) behaviour for $\varphi_m$ up to $\zeta \approx 20$ in the supercritical regime. The data for $\varphi_h$, however, are better approximated by a linear dependence in log-log coordinates with a slope slightly less than unity (not shown).

Comparison of the subcritical (Figs. 10b and 11b) and supercritical (Figs. 10c and 11c) cases shows that the data points overlap for a range of $\zeta$ values. For this reason, using $\zeta$ as a threshold to filter out non-Monin-Obukhov cases is not a useful approach because even a restriction $\zeta < 1$ imposed on the data will allow data with non-Kolmogorov turbulence to pass.

We found that prerequisites (10a) and (10b) play different roles if they are imposed on the data separately. Figures 12 shows plots of $\varphi_m$ versus $\zeta$ when a single prerequisite (10a) with $Ri_{cr} = 0.2$ (Fig. 12a) and a single prerequisite (10b) with $Rf_{cr} = 0.2$ (Fig. 12b) has been imposed on the data. According to Fig. 12a, the use of the single threshold $Ri_{cr} = 0.2$ does not improve the situation much as compared with the original plot in Figs. 10a. Even setting a tighter restriction, $Ri_{cr} = 0.15$, improves the situation only slightly more but does not eliminate a deviation from the $z$-less predictions (not shown). Meanwhile, Fig. 12b shows that implementing the single prerequisite (10b) with $Rf_{cr} = 0.2$ leads to results that are closely similar to those shown in Figs. 10b, where both restrictions (10a) and (10b) were used. Similar results have been also obtained for $\varphi_h$ (not shown). Thus, a primary threshold is associated with the critical flux Richardson number $Rf_{cr}$ (see discussion below).

Furthermore, we found that the choice of $Rf_{cr}$ affects the value of $\beta_m$. Figure 13 shows plots of $\varphi_m$ and $\varphi_h$ versus $\zeta$ when both prerequisites (10) with $Ri_{cr} = Rf_{cr} = 0.25$ have been imposed on the data. These results are pretty similar to those shown in Figs. 10b and 11b for $Ri_{cr} = Rf_{cr} = 0.2$, noting that prerequisite (10a) plays a minor role (cf. Fig. 12).

According to Fig. 13, the data better fit the linear equations (9) when both numerical coefficients, $\beta_m$ and $\beta_h$, equal 4 (blue dashed-dotted line) rather than 5 (black dashed lines), as in Figs. 10b and 11b. However, this result refers more to $\varphi_m$ (Fig. 13a) than to $\varphi_h$ (Fig. 13b). As discussed in Sect. 4, it is believed that this is due to the fact that $Rf_{cr} = 1/\beta_m$ (see Yamada 1975; Garratt 1992; Wyngaard 2010; Zilitinkevich et al. 2010), which in combination with the inequality (10b) leads to the observed effect. Note that, in the plot of $\varphi_m$ versus $\zeta$, prerequisite (10b) excludes data points that are below the line $\varphi_m = \zeta / Rf_{cr}$ (similar to the dashed area in Figs. 1 and 3 of Baas et al. 2008). However, presumably, this effect is true only for $Rf_{cr}$ around the critical point 0.20-0.25. In other words, experimental data do not follow the linear dependence (9) with $\beta_m = 2$ when prerequisite (10b) with $Rf_{cr} = 0.5$ is imposed on the data (not shown).

The criteria in Eqs. 10a, 10b can be also applied to the dimensionless standard deviations (or variances) of wind-speed components and the sonic temperature (4). We consider the standard deviation of the vertical wind speed component in detail and briefly the standard deviations for the *u* and *v* velocity components and the sonic temperature. Figure 14 shows plots as in Figs. 10 and 11 but for the normalized standard deviation of the vertical wind speed component, $\varphi_w = \sigma_w / u_*$ (local scaling). According to Fig. 14a, the original data do not follow the *z*-less prediction because they monotonically increase with increasing $\zeta$ (cf. Pahlow et al. 2001). However, the universal function $\varphi_w$ becomes asymptotically constant when both prerequisites (10a) and (10b) with $Ri_{cr} = Rf_{cr} = 0.20$ are imposed on the data; that is, the SHEBA data for $\sigma_w / u_*$ are consistent with the classical Monin-Obukhov *z*-less prediction (Fig. 14b) after the irrelevant cases (turbulence without the Richardson-Kolmogorov cascade) have been filtered out.

In the supercritical regime (Fig. 14c), $\varphi_w$ is approximately constant in the range $0.1 < \zeta < 3$ (*z*-less behaviour). The bias for level 1 and partially for level 2 in Fig. 14 is due to the close proximity of these levels to the surface, as discussed earlier. Overall, the behaviour of $\varphi_w = \sigma_w / u_*$ for different prerequisites (original dataset, subcritical and supercritical regimes) is consistent with the patterns for $\varphi_m$ and $\varphi_h$ (Figs. 10 and 11). The horizontal dashed lines in Fig. 14 correspond to $\varphi_w = \sigma_w / u_* = 1.33$, which is a median value computed in the subcritical regime (Fig. 14b).

By analogy with Fig. 12, we tested plots of $\varphi_w = \sigma_w / u_*$ versus $\zeta$ when prerequisites (10a and 10b) were imposed on the data separately (Fig.15). According to Fig. 15, the results are similar to those shown in Figs. 12 for $\varphi_m$; that is, implementing a single prerequisite (10a) with $Ri_{cr} = 0.2$ does not improve the situation (Fig. 15a) as compared with the original dataset (Fig.

14a). In contrast, $Rf_{cr} = 0.2$ can be considered a primary threshold for $\sigma_w/u_*$ (Fig. 15b). Similar results have been also obtained for other standard deviations (not shown).

One may suggest several reasons why $Rf_{cr}$, but not $Ri_{cr}$, is a primary threshold for MOST applicability. First, there are fundamental grounds. As mentioned in Sect. 4, a transition from turbulent flow to laminar (generally non-turbulent) flow is described by $Rf_{cr}$ that follows from the TKE budget equation. The reverse laminar-turbulent transition is associated with $Ri_{cr}$. Second, the majority of points in our plot of $Ri$ versus $Rf$ (Fig. 9) is located below the diagonal line $Ri = Rf$ (see Sect. 4), and the threshold $Rf_{cr} = 0.2$ will filter out more irrelevant points than the threshold $Ri_{cr} = 0.2$. Another reason is associated with the structure of the relationship (6), which in combination with the inequality (10b) imposes constraints on $\varphi_m$ and $\varphi_h$ (e.g., Fig. 13); but this point is more relevant to the flux-profile relationships (3) than to the flux-variance relationships (4).

Analysis of the presumably universal functions (3) and (4) presented in Figs. 10-15 generally proves the validity of our approach ("return to $z$-less" by separation of the data points into subcritical and supercritical cases, i.e., "separating the apples from the oranges"). However, all these plots are affected by self-correlation because $u_*$ appears both in the definitions of the universal functions and in $\zeta$ (e.g. Grachev et al. 2007b, 2012 and references therein). Furthermore as mentioned above, $z$-less behaviour of $\varphi_m$ in Fig. 10b is substantially associated with the prerequisite (10b), which excludes data points below the line $\varphi_m = \beta_m \zeta$ in the plot of $\varphi_m$ versus $\zeta$ where $\beta_m = 1/Rf_{cr}$.

Here we propose a method that does not contain these flaws. This method is based on the idea that a combination of any universal functions is itself a universal function for which the $z$-less scaling should be also valid. For example, the function $\varphi_m \varphi_w^{-1} = \left(\dfrac{\kappa z}{\sigma_w}\right)\dfrac{dU}{dz}$ plotted versus $\zeta$ by definition is not affected by the self-correlation because a "new" universal function shares no variables with $\zeta$ except a reference height $z$ (formally $u_*$ in Eq. 3a is replaced by $\sigma_w$). Other examples of such functions are $\varphi_\varepsilon \varphi_w^{-3}$ and $\varphi_\varepsilon \varphi_m^{-3}$, where $\varphi_\varepsilon$ is the non-dimensional dissipation rate of TKE (e.g. Kaimal and Finnigan 1994).

Figure 16 shows plots as in Figs. 10, 11, and 14 but for the universal function $\varphi_m \varphi_w^{-1}$. According to Fig. 16a, the original data increase more slowly with increasing $\zeta$ than is predicted by $z$-less scaling. However, according to Figs. 16b, the function is consistent with the $z$-less prediction, $\varphi_m \varphi_w^{-1} \propto \zeta$, when both prerequisites (10) with $Ri_{cr} = Rf_{cr} = 0.20$ have been imposed on the data. Thus, as mentioned above, this result is not affected by the self-correlation and is a critical test for our approach. Note that this method can also be used in the free-convection limit where self-correlation is also a serious issue.

Although the inequalities (10) do not impose restrictions on $\zeta$ (mathematically $\zeta$ can become indefinitely large), in effect, implementing the prerequisite (10) removes most of the data with $\zeta > 6$ in the case $Ri_{cr} = Rf_{cr} = 0.2$ (Figs. 10b, 11b, and 14b) and with $\zeta > 30$ in the case $Ri_{cr} = Rf_{cr} = 0.25$ (Fig. 13). At the same time, in the supercritical regime, $\zeta$ can reach values as low as $\zeta \approx 0.1$ and less (Figs. 10c, 11c, and 14c).

**6 Discussion and Concluding Remarks**

The applicability of classical similarity theory (Monin and Obukhov 1954) has been limited by the assumption of a constant-flux layer, which is valid in a narrow range $\zeta < 0.1$ for stable conditions (e.g., Grachev et al. 2005, their Fig. 2). Nieuwstadt (1984) extended the range of applicability of the original theory for the SBL using local scaling (height-dependent fluxes) in place of surface scaling, but the limits of applicability of local similarity theory have been blurry.

In this study, we derive the limits of applicability of local similarity theory in the SBL based on the spectral analysis of wind velocity and temperature fluctuations. Measurements of small-scale atmospheric turbulence made over the Arctic pack ice during the Surface Heat Budget of the Arctic Ocean experiment (SHEBA) are used to consider this problem. Our approach is based on the notion that the region of applicability of similarity theory for the flux-profile relationships (3) and for any other properly scaled statistics of the turbulence (Monin and Obukhov 1954) is the same as for the universal spectra and cospectra (Kaimal et al. 1972; Kaimal 1973). This idea arises because the turbulent fluxes and the variances are derived through frequency integration of the appropriate cospectra and spectra in the high-frequency range associated with energy-containing/flux-carrying eddies. Kolmogorov's theory describes the energy cascade and assumes that energy is transferred from larger eddies to successively smaller eddies – from the energy containing range to the dissipation range through the inertial subrange. Therefore, a collapse of the energy-containing turbulence causes the inertial subrange to collapse, and Kaimal's spectral and cospectral similarity becomes invalid. This collapse, in turn, leads to violations of flux-profile and flux-variance similarity. Thus, the upper limit of MOST in the SBL coincides with the region for which the –5/3 Kolmogorov power law is applicable.

We found that the −5/3 power law in the inertial subrange for stable conditions is obeyed over a wide range of the stability parameters; but when both the gradient Richardson number (5) and the flux Richardson number (6) exceed a 'critical value' of about 0.20-0.25, the inertial subrange associated with a Richardson-Kolmogorov energy cascade dies out (Figs. 7 and 8). Thus, according to our data, the upper limit of applicability for the −5/3 Kolmogorov power law (the Obukhov-Corrsin law for the passive scalar) and, therefore, for the existence of the inertial subrange in the SBL, is described by inequalities (10) with $Ri_{cr} = Rf_{cr} = 0.20 - 0.25$. As argued above, the inequalities (10) also describe the applicability limit for local similarity theory in the SBL.

However, we found that prerequisites (10a) and (10b) play different roles if they are imposed on the data separately. The primary threshold is associated with the critical flux Richardson number $Rf_{cr}$. That is, the prerequisite (10b) may be used alone without (10a), but (10a) cannot be used alone. Apparently this constraint is associated with the fact that the majority of the data points in our plot of $Ri$ versus $Rf$ (Fig. 9) is located in the region $Ri < Rf$.

We emphasise that, although $\zeta$ is a primary scaling parameter in the MOST, the limit of applicability of local similarity theory does not depend on $\zeta$. Note that our approach imposes tighter restrictions on the flux Richardson number, $Rf < Rf_{cr} = 0.20 - 0.25$, than does the traditionally used restriction $Rf < 1$, which is simply that shear production is larger than buoyancy destruction (e.g., Baas et al. 2008, their Figs. 1 and 3 and references therein). In the framework of MOST, only one threshold should be used ($Ri$ or $Rf$); moreover, neither $Ri$ nor $Rf$ has an advantage over the other. For example, using $\beta_m = \beta_h = 5$ leads to $Ri_{cr} = Rf_{cr} = 0.2$ (see Eqs. 5, 6, and 8). In practice, because the experimental data scatter, both inequalities (10) should be used to filter out irrelevant cases (see Fig. 9); and, herewith, $Rf_{cr}$ is the primary threshold.

The 'critical' Richardson number 0.20 – 0.25 obtained here should be not confused with the classical critical Richardson number beyond which turbulence cannot be sustained. According to our Figs. 5 and 6, low levels small-scale turbulence survives even in the supercritical regime. This result is consistent with other data (see Sect. 4). However, this small-scale 'supercritical' turbulence is non-Kolmogorov turbulence (more accurately 'turbulence without Richardson-Kolmogorov cascade'), intermittent, and non-stationary; and it decays rapidly with further increasing stability. While the flux data do not reveal an obvious critical Richardson number, Kolmogorov's five-thirds law fails when $Ri$ and $Rf$ exceed a critical value around 0.20 – 0.25. Thus, $Ri_{cr} = Rf_{cr} = 0.20 - 0.25$ can be considered as a point when turbulence properties change, for example, as a phase transition. Thus, in the canonical sense, the critical Richardson number for the suppression of turbulence does not exist because turbulence never disappears for $Ri > 0.20 - 0.25$ and $Rf > 0.20 - 0.25$ and there is no evidence that the suppression of turbulence by stable stratification causes a transition to laminar flow. However, $Ri_{cr} = Rf_{cr} = 0.20 - 0.25$ is certainly a critical point because it separates turbulence with completely different properties.

This view is consistent with the laboratory experiments of the Van Atta group (Rohr et al. 1988) and the model results of Baumert and Peters (2004). It is interesting that the values obtained here coincide with the canonical values of 0.20 or 0.25 for the 'classical' critical Richardson number but were derived from quite different reasoning (e.g., by Miles 1961; Howard 1961). It remains unknown whether this is simply a coincidence.

Applying the prerequisite (10) shows that the data follow classical Monin-Obukhov local $z$-less predictions after the irrelevant cases (turbulence without the Richardson-Kolmogorov cascade) have been filtered out (Figs. 10b, 11b, and 14b). Thus, our approach removes the controversy associated with the $z$-less concept mentioned in Sect. 2. Furthermore, the decaying turbulence in the supercritical regime also shows $z$-less behaviour within a certain range of stable stratification (Figs. 10c, 11c, and 14c). Note that the 'supercritical turbulence' is strongly affected by the turning effects of the Coriolis force, as found earlier by Grachev et al. (2005, 2008).

Frequently, the SBL has been classified into several separate regimes. Mahrt et al. (1998) divided turbulence into weakly stable, transition stability, and very stable regimes based on $z/L$. Mauritsen and Svensson (2007) and Zilitinkevich et al. (2007) proposed a similar classification, but they argued in favour of the gradient Richardson number (5) as a basic stability parameter. Our study allows for the refining of this scheme.

Our analysis shows that, first, the traditional SBL can be classified into two major regimes: subcritical and supercritical, which are separated on the basis of the criteria (10a) and (10b). The subcritical regime is associated with Kolmogorov turbulence, and different turbulent statistics can be described by similarity theory. This regime, in its turn, can be divided into the surface-layer regime, which is governed by original MOST predictions ($\zeta < 0.1$ or $Ri < 0.1$ or $Rf < 0.1$), and the local $z$-less regime ($0.1 < Ri < 0.20 - 0.25$ and $0.1 < Rf < 0.20 - 0.25$). This division corresponds to the weakly stable boundary layer and the transition regime of Mahrt et al. (1998) and Mauritsen and Svensson (2007). At the high-end of the subcritical regime, the Coriolis effect may be relevant (Grachev et al. 2005, 2008).

The supercritical stable regime (or very stable case) is associated with non-Kolmogorov, non-stationary (decaying) turbulence and the strong influence of the earth's rotation, even near the surface. Traditional similarity theory appears to break down in this regime. However, in the subrange $0.20 - 0.25 < Ri < O(0.5)$ and $0.20 - 0.25 < Rf < O(0.5)$, some small-scale turbulence survives passing through the critical point; and, presumably, the $z$-less concept (8) may also be applied in this subrange but not in the framework of MOST. Beyond the point $Ri \approx O(0.5)$ and $Rf \approx O(0.5)$, apparently, shear production is too small to maintain small-scale turbulence; and the turbulence, if it exists, is generated or modulated by different instability mechanisms and external events in the low-frequency spectral range. Our classification (especially for $Ri > 0.25$) is consistent with the SBL regimes suggested by Baumert and Peters (2009).

Thus, our approach is based on the separation of data points into subcritical and supercritical cases, i.e., "separating the apples from the oranges". To evaluate different MOST functions, we filtered out all data points where the Richardson-Kolmogorov cascade is not observed by imposing the prerequisite (10) on the data. Namely, this approach was used in the Kansas (1968) milestone experiment where the data with $Ri > 0.20 - 0.25$ were excluded from the analysis (e.g., Businger et al. 1971; Kaimal 1973). Obviously, the applicability limit (10) is relevant not only for the MOST (flux-based scaling) but also can be applied to $\sigma_w$ scaling (Sorbjan 2010) and to gradient-based scaling (Sorbjan 2008, 2010; Sorbjan and Grachev 2010), which is formally equivalent to the MOST approach.

**Acknowledgements**   The U.S. National Science Foundation's Office of Polar Programs supported our original SHEBA research. During the current work NSF also supported AAG and POGP with award ARC 11-07428 and ELA with award ARC 10-19322. AAG also was supported by the U.S. Civilian Research & Development Foundation (CRDF) with award RUG1-2976-ST-10.

Smedman A-S. (1988) Observations of a multi-level turbulence structure in a very stable atmospheric boundary layer. *Boundary-Layer Meteorol*. **44**: 231–253

Sorbjan Z. (1986) On similarity in the atmospheric boundary layer. *Boundary-Layer Meteorol*. **34**: 377–397

Sorbjan Z. (1988) Structure of the stably-stratified boundary layer during the SESAME-1979 experiment. *Boundary-Layer Meteorol*. **44**: 255–260

Sorbjan Z. (1989) *Structure of the atmospheric boundary layer*. Prentice-Hall, Englewood Cliffs, NJ, 317 pp

Sorbjan Z. (2008) Gradient-based similarity in the atmospheric boundary layer. *Acta Geophysica*. **56**(1): 220–233

Sorbjan Z. (2010) Gradient-based scales and similarity laws in the stable boundary layer. *Q. J. R. Meteorol. Soc.* **136**(650A): 1243–1254

Sorbjan, Z. Grachev, A.A. (2010) An evaluation of the flux-gradient relationship in the stable boundary layer. *Boundary-Layer Meteorol*. **135**(3): 385–405

Stull R.B. (1988) *An Introduction to boundary-layer meteorology*. Kluwer Academic Publishers, Boston, Mass., 666 pp

Townsend A. (1958) Turbulent flow in a stably stratified atmosphere. *J. Fluid Mech*. **3**: 361–372

Van de Wiel B.J.H., Moene A.F., Steeneveld G.J., Hartogensis O.K., Holtslag A.A.M. (2007) Predicting the collapse of turbulence in stably stratified boundary layers. *Flow Turbulence Combust*. 79:251–274, doi 10.1007/s10494-007-9094-2

Vickers D., Mahrt L. (1997) Quality control and flux sampling problems for tower and aircraft data. *J. Atmos. Oc. Tech.* **14**(3): 512–526
42

**Figure Captions**

Fig. 1. Typical raw energy spectra of (*a*) the longitudinal, (*b*) lateral, (*c*) vertical velocity components and (*d*) the sonic temperature at five levels for weakly and moderately stable conditions measured on 5 December 1997 (YD 339.375 UTC). Stability parameters generally increase with increasing height (levels from 1 to 5): $\zeta \approx 0.116, 0.210, 0.498, 1.23, 2.22$; $Ri \approx 0.282, 0.124, 0.108, 0.124, 0.152$; $Rf \approx 0.151, 0.102, 0.117, 0.155, 0.197$.

Fig. 2. Typical raw cospectra of (*a*) the downwind stress and (*b*) the sonic temperature flux at five levels for weakly and moderately stable conditions measured during the same time as in Fig. 1, 5 December 1997 (YD 339.375 UTC). For data presented here and in Fig. 1, the friction velocity and the sensible heat flux are approximately constant or decrease with height (levels from 1 to 5): $u_*$ (m s$^{-1}$) $\approx 0.142, 0.140, 0.115, 0.091, 0.080$; $H_S$ (W m$^{-2}$) $\approx -15.6, -17.4, -13.9, -9.62, -7.67$.

Fig. 3. Typical raw energy spectra of the (*a*) longitudinal, (*b*) lateral, and (*c*) vertical velocity components and (*d*) the sonic temperature at five levels for moderate and very stable conditions measured on 30 December 1997 (YD 364.0833 UTC). Stability parameters generally increase with increasing height (levels from 1 to 5): $\zeta \approx 1.71, 1.63, 2.59, 8.54, 38.6$; $Ri \approx 0.103, 0.118, 0.150, 0.210, 0.2844$; $Rf \approx 0.259, 0.225, 0.232, 0.289, 0.672$.

Fig. 4. Typical raw cospectra of (*a*) the downwind stress and (*b*) the sonic temperature flux at five levels for very stable conditions measured during the same time as in Fig. 3, 30 December 1997 (YD 364.0833 UTC). For data presented here and in Fig. 3, the friction velocity and the sensible heat flux monotonically decrease with height (levels from 1 to 5): $u_*$ (m s$^{-1}$) $\approx 0.0436, 0.0529, 0.0443, 0.0209, 0.0125$; $H_S$ (W m$^{-2}$) $\approx -6.94, -7.61, -4.27, -0.835, -0.513$.

Fig. 5. Behavior of the bin-averaged downwind momentum flux $\tau = -\rho <u'w'>$ (bin medians) in the vicinity of the critical Richardson number for five levels of the main SHEBA tower during the 11 months of measurements plotted versus (*a*) the gradient Richardson number (5), (*b*) the flux Richardson number (6), and (c) the Monin-Obukhov stability parameter (1) ($Rf$ and $z/L$ are based on local scaling). Individual 1-hr averaged SHEBA data based on the median fluxes for the five levels are shown as the background x-symbols. The vertical dashed lines correspond to $Ri$ and $Rf = 0.2$.

Fig. 6. Same as Fig. 5 but for the sensible heat flux $H_S = c_p \rho <w'\theta'>$.

Fig. 7. Plots of the bin-averaged spectral slope (bin medians) in the inertial subrange for the one-dimensional spectrum of the (a) longitudinal, (b) lateral, and (c) vertical velocity components and (d) the sonic temperature for five levels of the main SHEBA tower measured during the 11 months as functions of the gradient Richardson number (5). The spectral slopes were computed in the 0.96 – 2.95 Hz frequency band. Individual 1-hr averaged SHEBA data for level 3 are shown as the background x-symbols. The vertical dashed lines correspond to $Ri = 0.2$. The horizontal dashed lines represent the –5/3 Kolmogorov power law for the inertial subrange. Data with small magnitudes of the temperature gradients (less than 0.03 K m$^{-1}$) in plot (d) have been omitted to avoid large uncertainty in computing the temperature spectra.

Fig. 8. Same as Fig. 7 but plotted versus the flux Richardson number (6) ($Rf$ is based on local scaling). The vertical dashed lines correspond to $Rf = 0.2$.

Fig. 9. Bin-averaged gradient Richardson number (5) (bin medians) plotted versus the flux Richardson number (6) for the five levels of the main SHEBA tower during the 11 months of measurements plotted. The vertical and horizontal dashed lines correspond to $Rf$ and $Ri = 0.2$, respectively. The dashed-dotted line is 1:1. The yellow x-symbols are medians of all good 1-hr measurements on the tower.

Fig. 10. The bin-averaged non-dimensional vertical gradient of mean wind speed $\varphi_m$ for five levels of the main SHEBA tower during the 11 months of measurements are plotted versus the Monin-Obukhov stability parameter for local scaling (a) for the original data, (b) in the subcritical regime when both prerequisites (10) with $Ri_{cr} = Rf_{cr} = 0.2$ have been imposed on the data, (*c*) in the supercritical regime when a prerequisite $Ri > Ri_{cr} = 0.2$ and $Rf > Rf_{cr} = 0.2$ has been imposed on the data. Individual 1-hr averaged SHEBA data based on the median fluxes for the five levels are shown as the background x-symbols.

Fig. 11. Same as Fig. 10 but for the non-dimensional vertical gradient of mean potential temperature $\varphi_h$.

Fig. 12. The bin-averaged non-dimensional vertical gradient of mean wind speed $\varphi_m$ for five levels of the main SHEBA tower during the 11 months of measurements are plotted versus the local Monin-Obukhov stability parameter when (*a*) a single prerequisite (10a) with $Ri_{cr} = 0.2$ and (*b*) a single prerequisite (10b) with $Rf_{cr} = 0.2$ has been imposed on the data.

Fig. 13. The bin-averaged non-dimensional vertical gradients of (*a*) mean wind speed $\varphi_m$ and (*b*) mean potential temperature $\varphi_h$ for five levels of the main SHEBA tower during the 11 months of measurements are plotted versus the local Monin-Obukhov stability parameter in the subcritical regime when both prerequisites (10) with $Ri_{cr} = Rf_{cr} = 0.25$ has been imposed on the data.

Fig. 14. Same as Fig. 10 but for the normalized standard deviation of the vertical wind speed component $\varphi_w = \sigma_w / u_*$ (local scaling). The horizontal dashed lines correspond to $\varphi_w = 1.33$.

Fig. 15. Same as Fig. 12 but for the normalized standard deviation of the vertical wind speed component $\varphi_w = \sigma_w / u_*$. The horizontal dashed lines correspond to $\varphi_w = 1.33$.

Fig. 16. Same as Fig. 10 but for the function $\varphi_m \varphi_w^{-1} = \left( \dfrac{\kappa z}{\sigma_w} \right) \dfrac{dU}{dz}$ which is a combination of the universal functions (3a) and (4a) for $\alpha = w$ and is not affected by the self-correlation. Symbols and notation are the same as in Fig. 10.

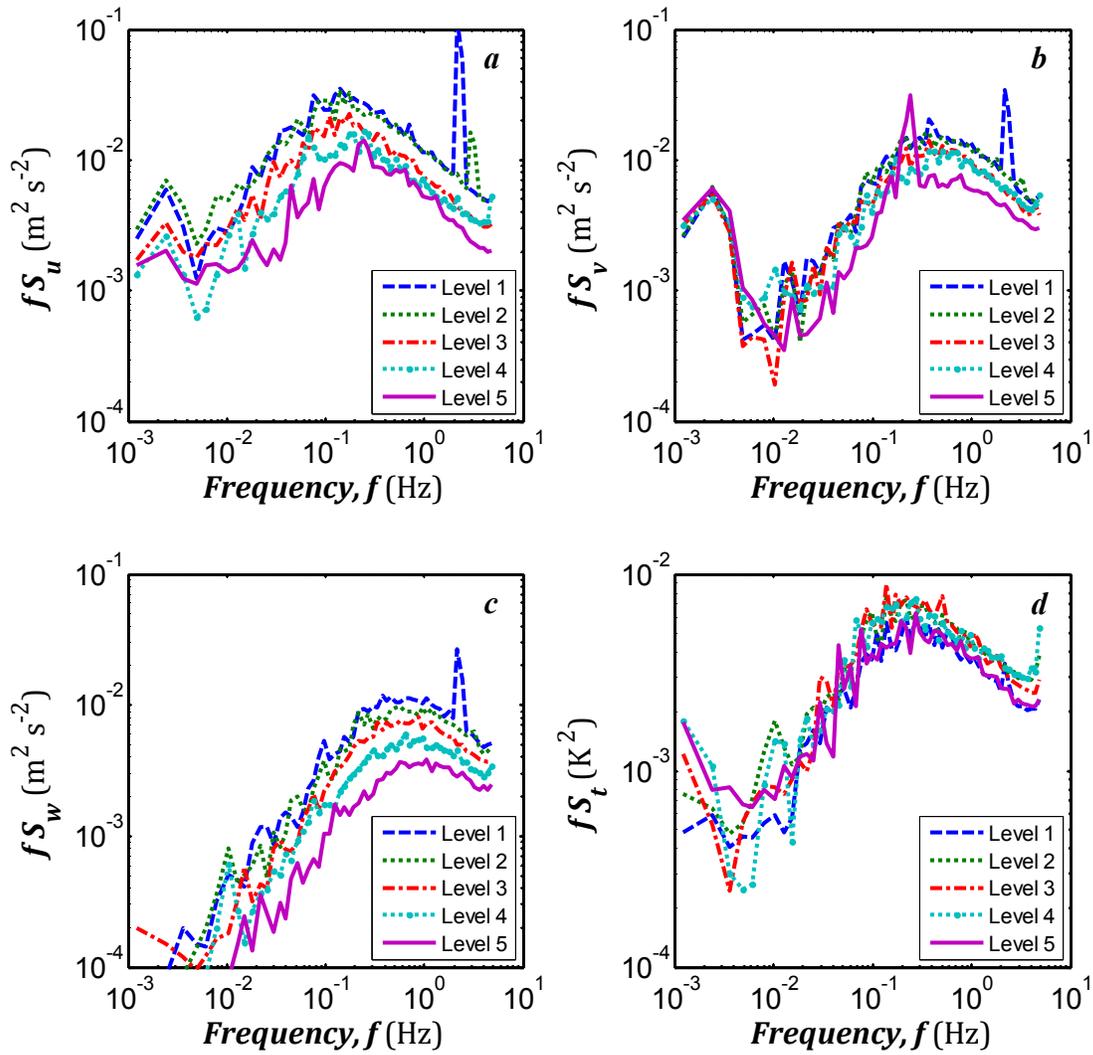

Fig. 1. Typical raw energy spectra of (*a*) the longitudinal, (*b*) lateral, (*c*) vertical velocity components and (*d*) the sonic temperature at five levels for weakly and moderately stable conditions measured on 5 December 1997 (YD 339.375 UTC). Stability parameters generally increase with increasing height (levels from 1 to 5): $\zeta \approx$ 0.116, 0.210, 0.498, 1.23, 2.22; $Ri \approx$ 0.282, 0.124, 0.108, 0.124, 0.152; $Rf \approx$ 0.151, 0.102, 0.117, 0.155, 0.197.

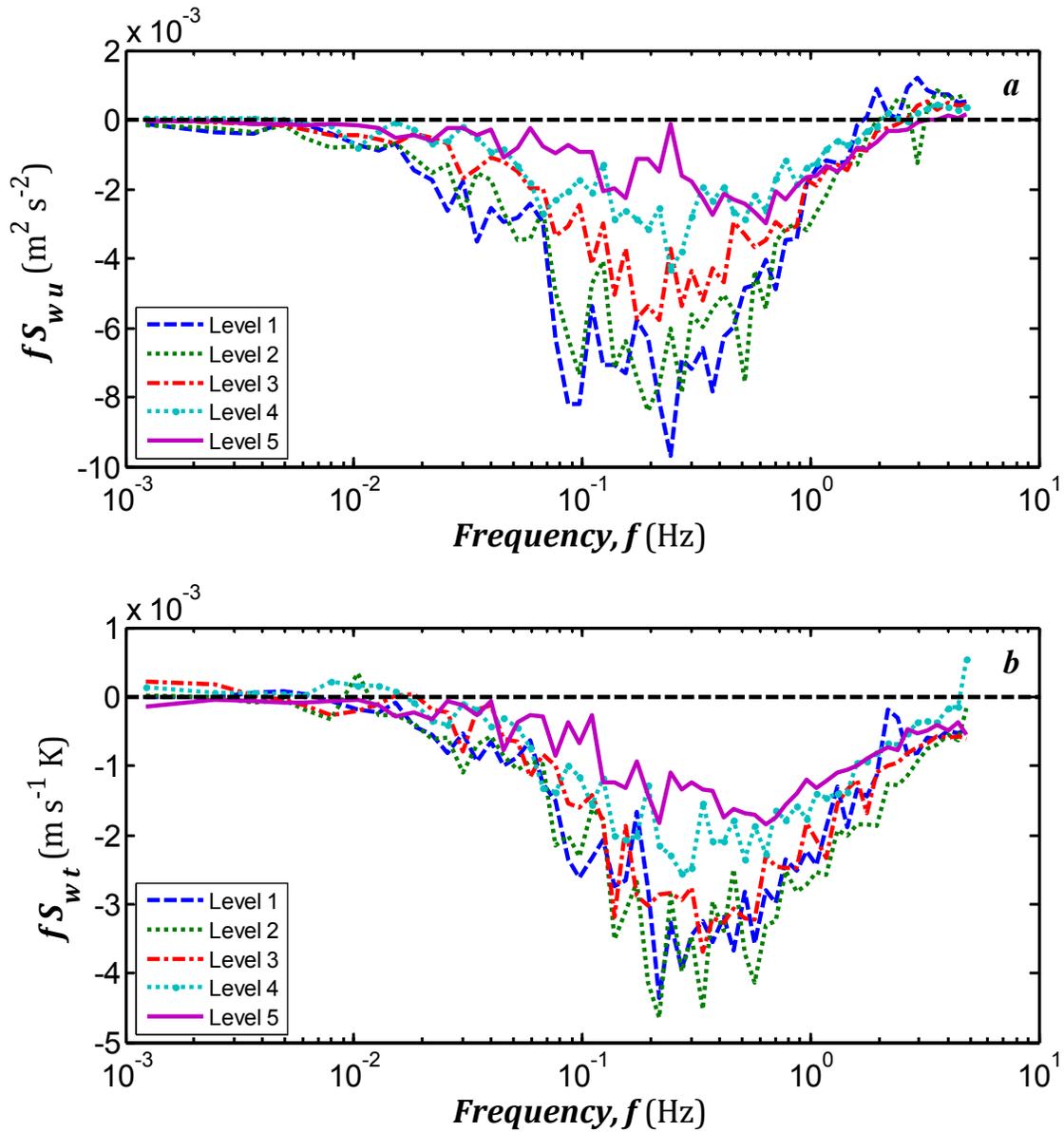

Fig. 2. Typical raw cospectra of (*a*) the downwind stress and (*b*) the sonic temperature flux at five levels for weakly and moderately stable conditions measured during the same time as in Fig. 1, 5 December 1997 (YD 339.375 UTC). For data presented here and in Fig. 1, the friction velocity and the sensible heat flux are approximately constant or decrease with height (levels from 1 to 5): $u_*$ (m s$^{-1}$) ≈ 0.142, 0.140, 0.115, 0.091, 0.080; $H_S$ (W m$^{-2}$) ≈ −15.6, −17.4, −13.9, −9.62, −7.67.

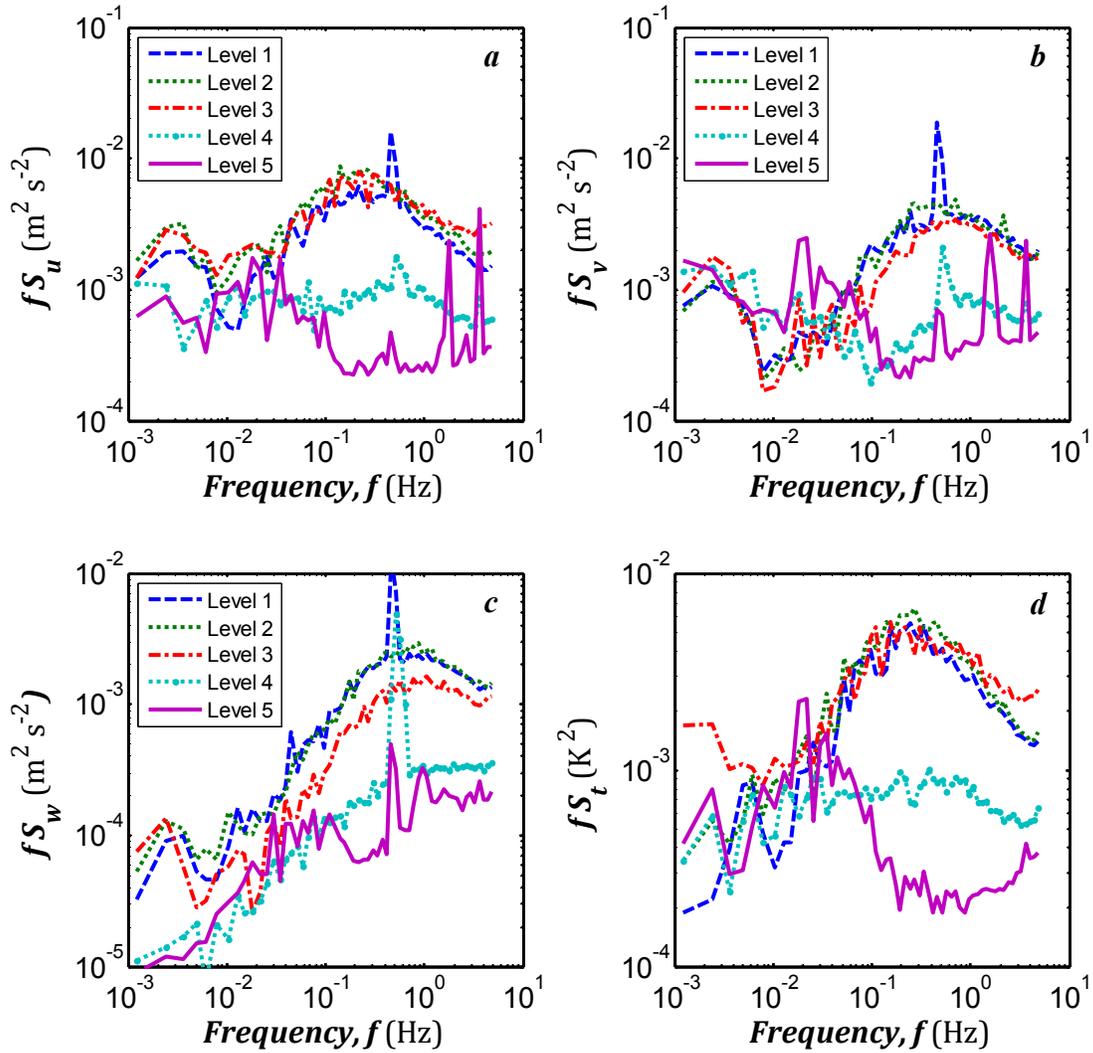

Fig. 3. Typical raw energy spectra of the (*a*) longitudinal, (*b*) lateral, and (*c*) vertical velocity components and (*d*) the sonic temperature at five levels for moderate and very stable conditions measured on 30 December 1997 (YD 364.0833 UTC). Stability parameters generally increase with increasing height (levels from 1 to 5): $\zeta \approx$ 1.71, 1.63, 2.59, 8.54, 38.6; $Ri \approx$ 0.103, 0.118, 0.150, 0.210, 0.2844; $Rf \approx$ 0.259, 0.225, 0.232, 0.289, 0.672.

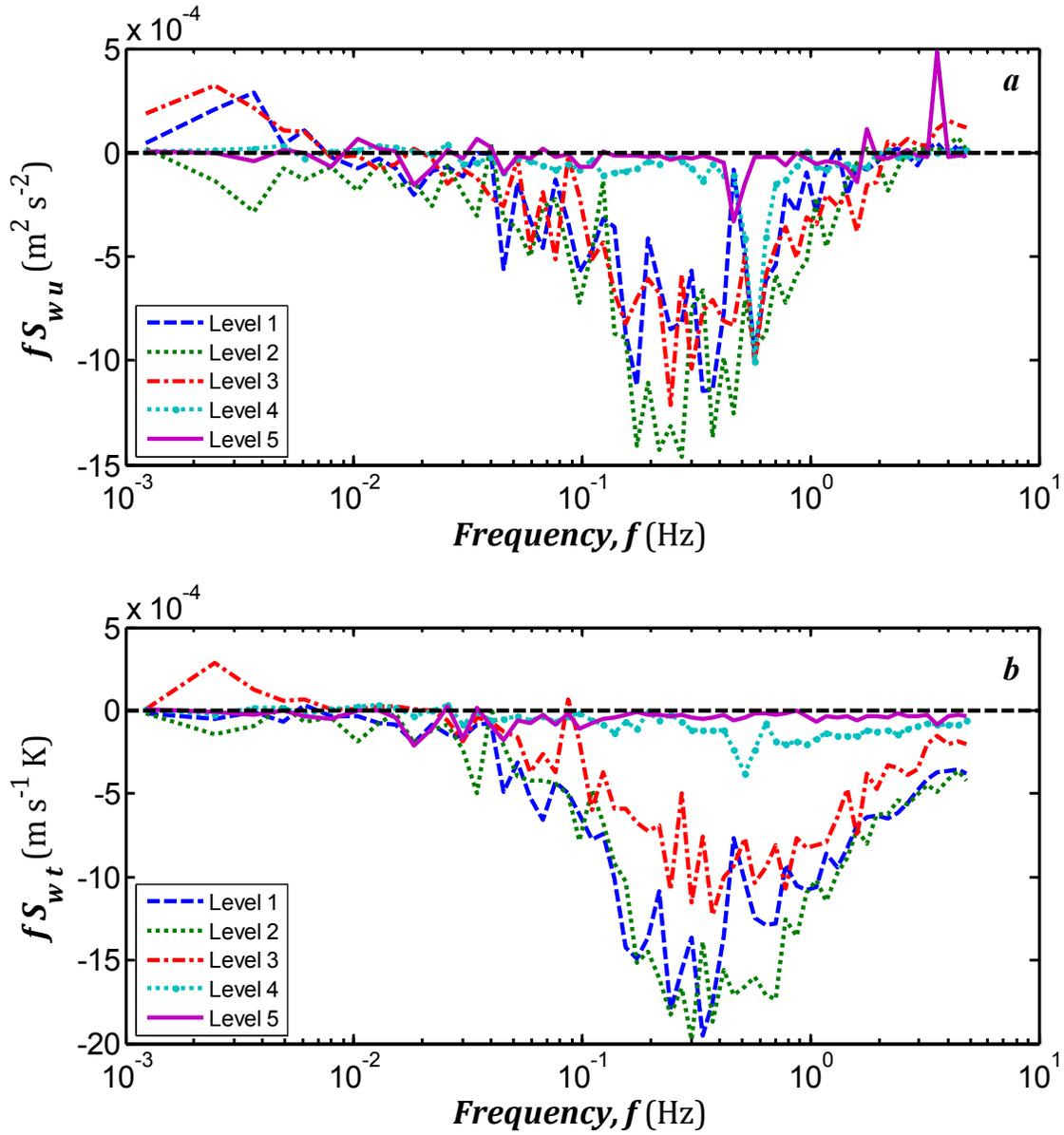

Fig. 4. Typical raw cospectra of (*a*) the downwind stress and (*b*) the sonic temperature flux at five levels for very stable conditions measured during the same time as in Fig. 3, 30 December 1997 (YD 364.0833 UTC). For data presented here and in Fig. 3, the friction velocity and the sensible heat flux monotonically decrease with height (levels from 1 to 5): $u_*$ (m s$^{-1}$) ≈ 0.0436, 0.0529, 0.0443, 0.0209, 0.0125; $H_S$ (W m$^{-2}$) ≈ –6.94, –7.61, –4.27, –0.835, –0.513.

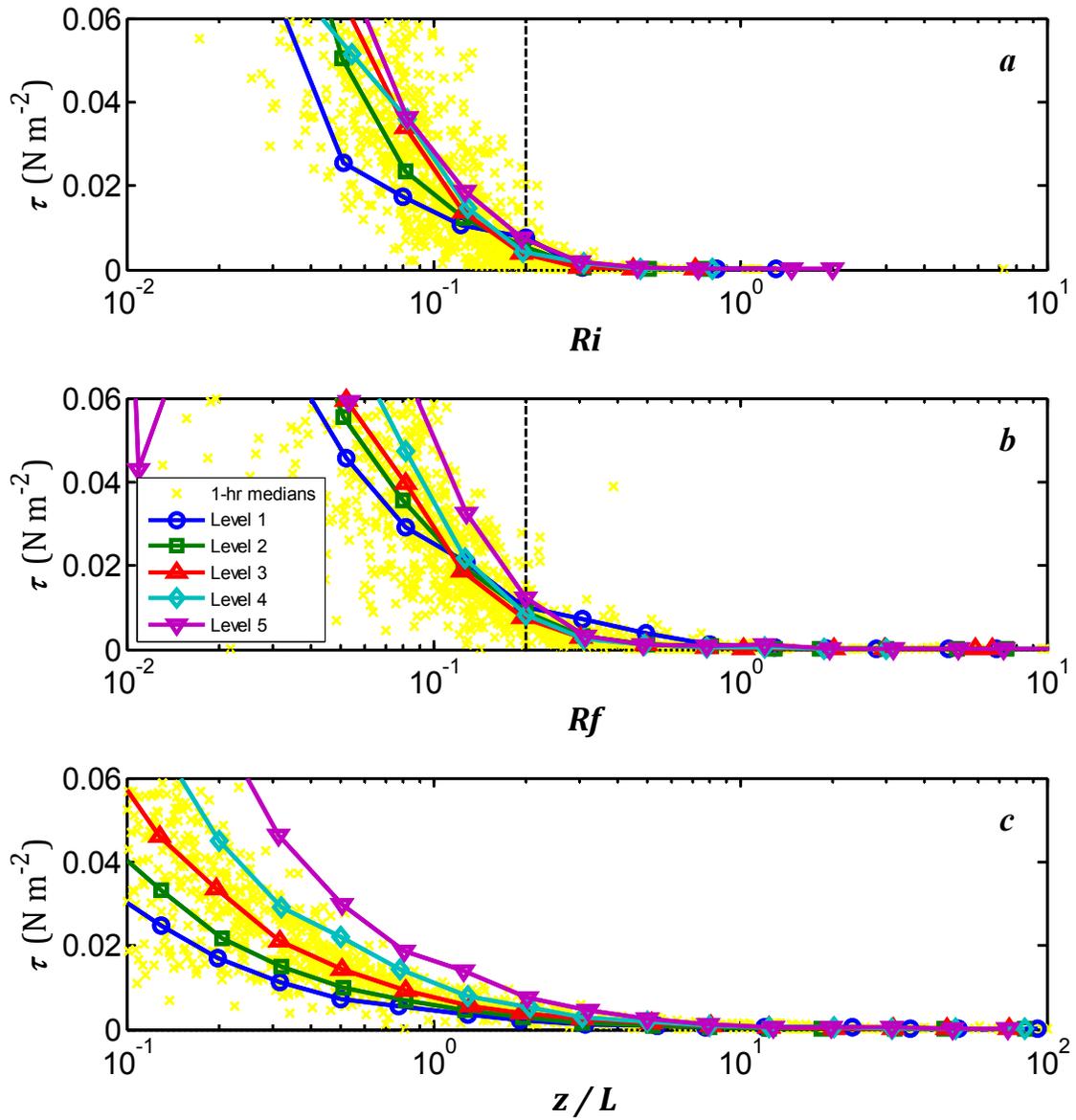

Fig. 5. Behavior of the bin-averaged downwind momentum flux $\tau = -\rho <u'w'>$ (bin medians) in the vicinity of the critical Richardson number for five levels of the main SHEBA tower during the 11 months of measurements plotted versus (*a*) the gradient Richardson number (5), (*b*) the flux Richardson number (6), and (*c*) the Monin-Obukhov stability parameter (1) (*Rf* and *z/L* are based on local scaling). Individual 1-hr averaged SHEBA data based on the median fluxes for the five levels are shown as the background x-symbols. The vertical dashed lines correspond to *Ri* and *Rf* = 0.2.

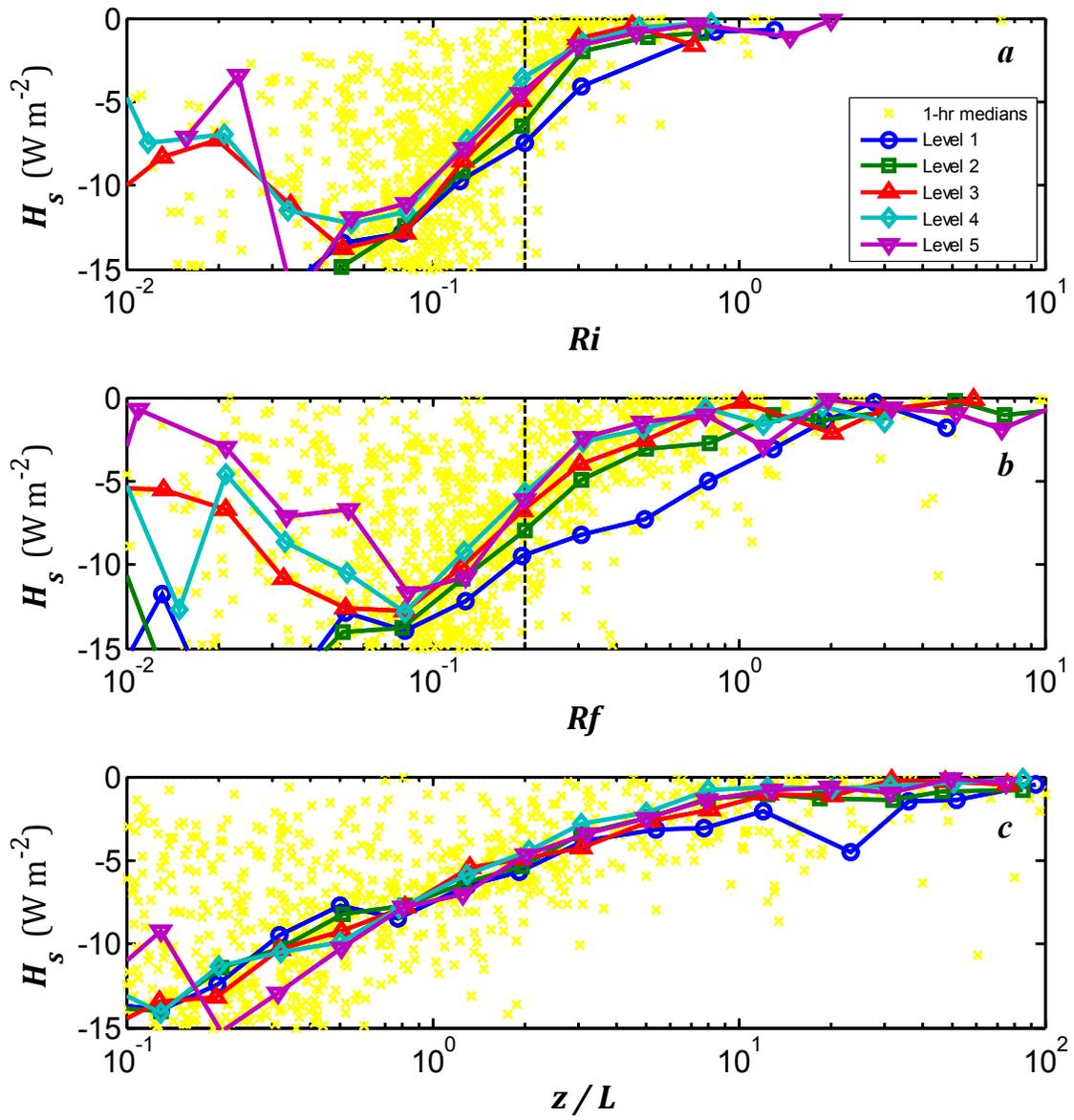

Fig. 6. Same as Fig. 5 but for the sensible heat flux $H_S = c_p \rho <w'\theta'>$.

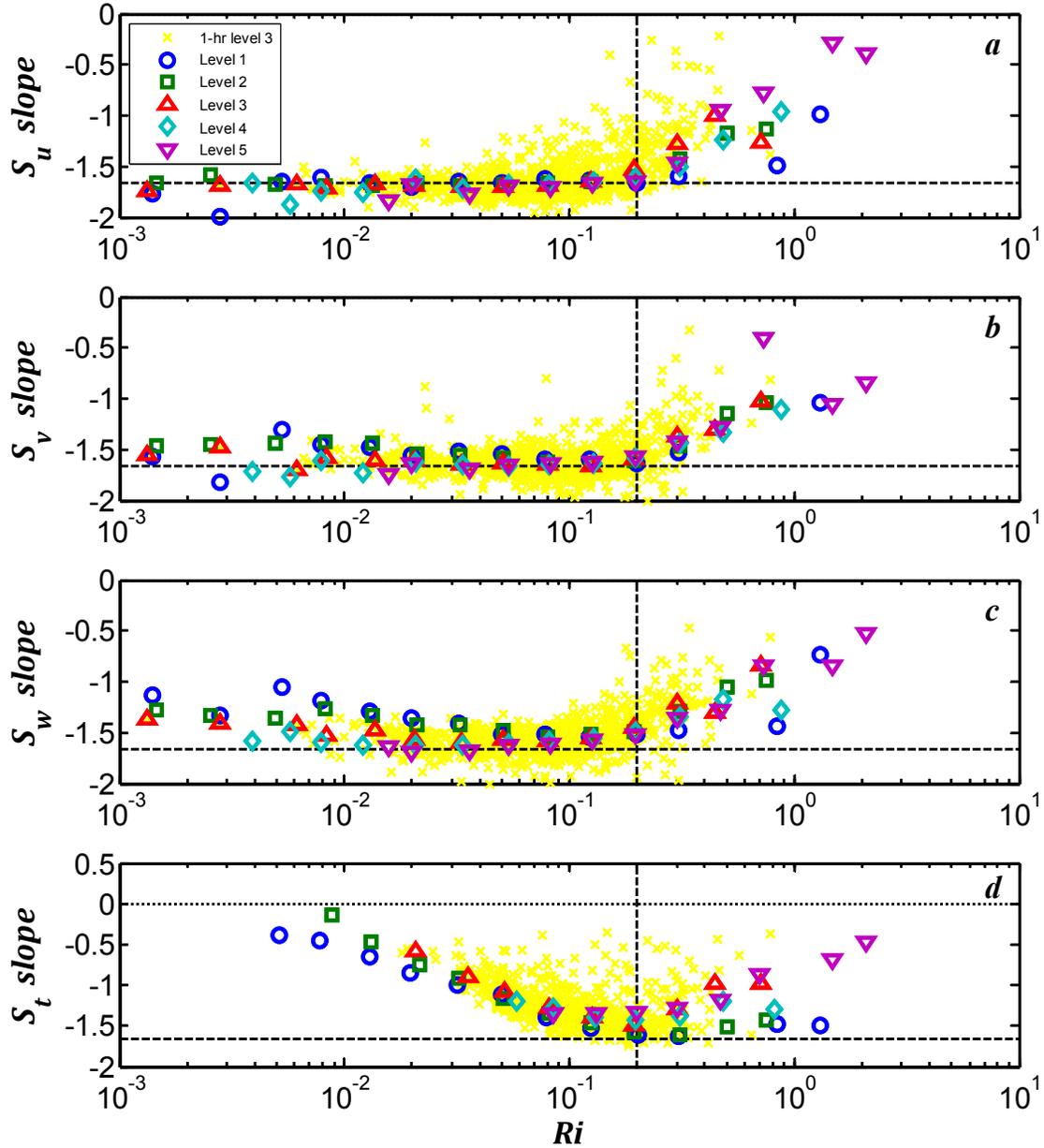

Fig. 7. Plots of the bin-averaged spectral slope (bin medians) in the inertial subrange for the one-dimensional spectrum of the (*a*) longitudinal, (*b*) lateral, and (*c*) vertical velocity components and (*d*) the sonic temperature for five levels of the main SHEBA tower measured during the 11 months as functions of the gradient Richardson number (5). The spectral slopes were computed in the 0.96 – 2.95 Hz frequency band. Individual 1-hr averaged SHEBA data for level 3 are shown as the background x-symbols. The vertical dashed lines correspond to $Ri$ = 0.2. The horizontal dashed lines represent the –5/3 Kolmogorov power law for the inertial subrange. Data with small magnitudes of the temperature gradients (less than 0.03 K m$^{-1}$) in plot (*d*) have been omitted to avoid large uncertainty in computing the temperature spectra.

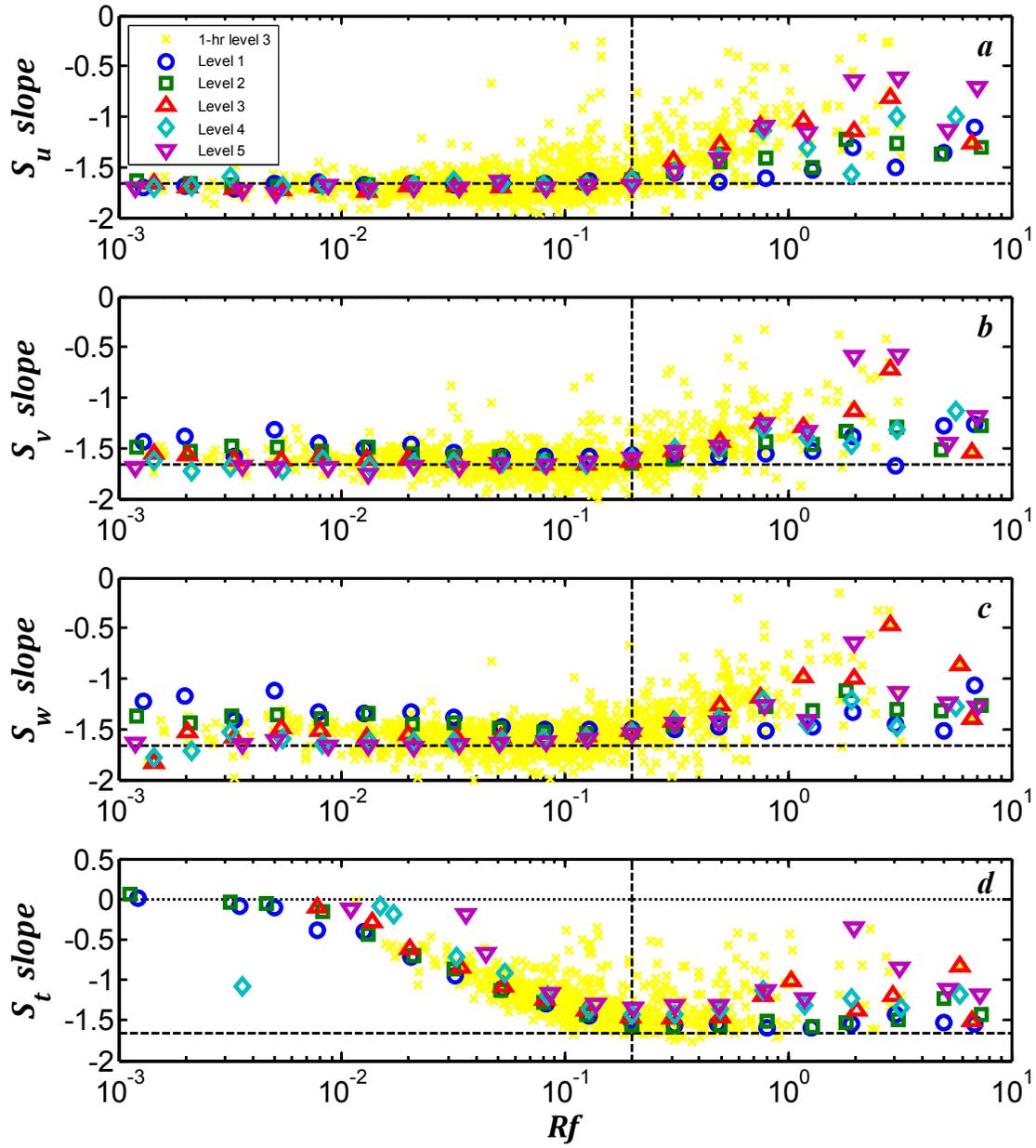

Fig. 8. Same as Fig. 7 but plotted versus the flux Richardson number (6) (*Rf* is based on local scaling). The vertical dashed lines correspond to *Rf* = 0.2.

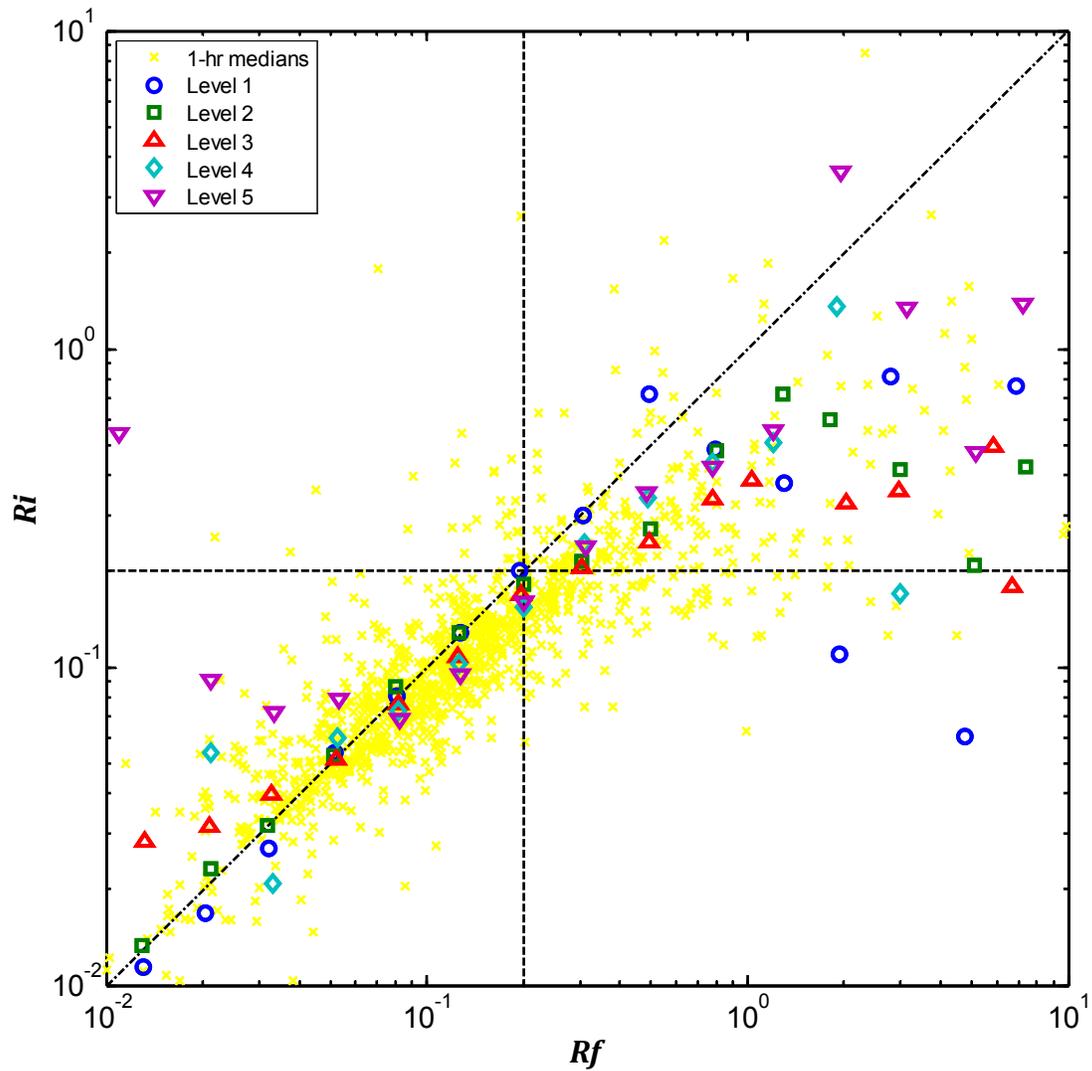

Fig. 9. Bin-averaged gradient Richardson number (5) (bin medians) plotted versus the flux Richardson number (6) for the five levels of the main SHEBA tower during the 11 months of measurements plotted. The vertical and horizontal dashed lines correspond to $Rf$ and $Ri = 0.2$, respectively. The dashed-dotted line is 1:1. The yellow x-symbols are medians of all good 1-hr measurements on the tower.

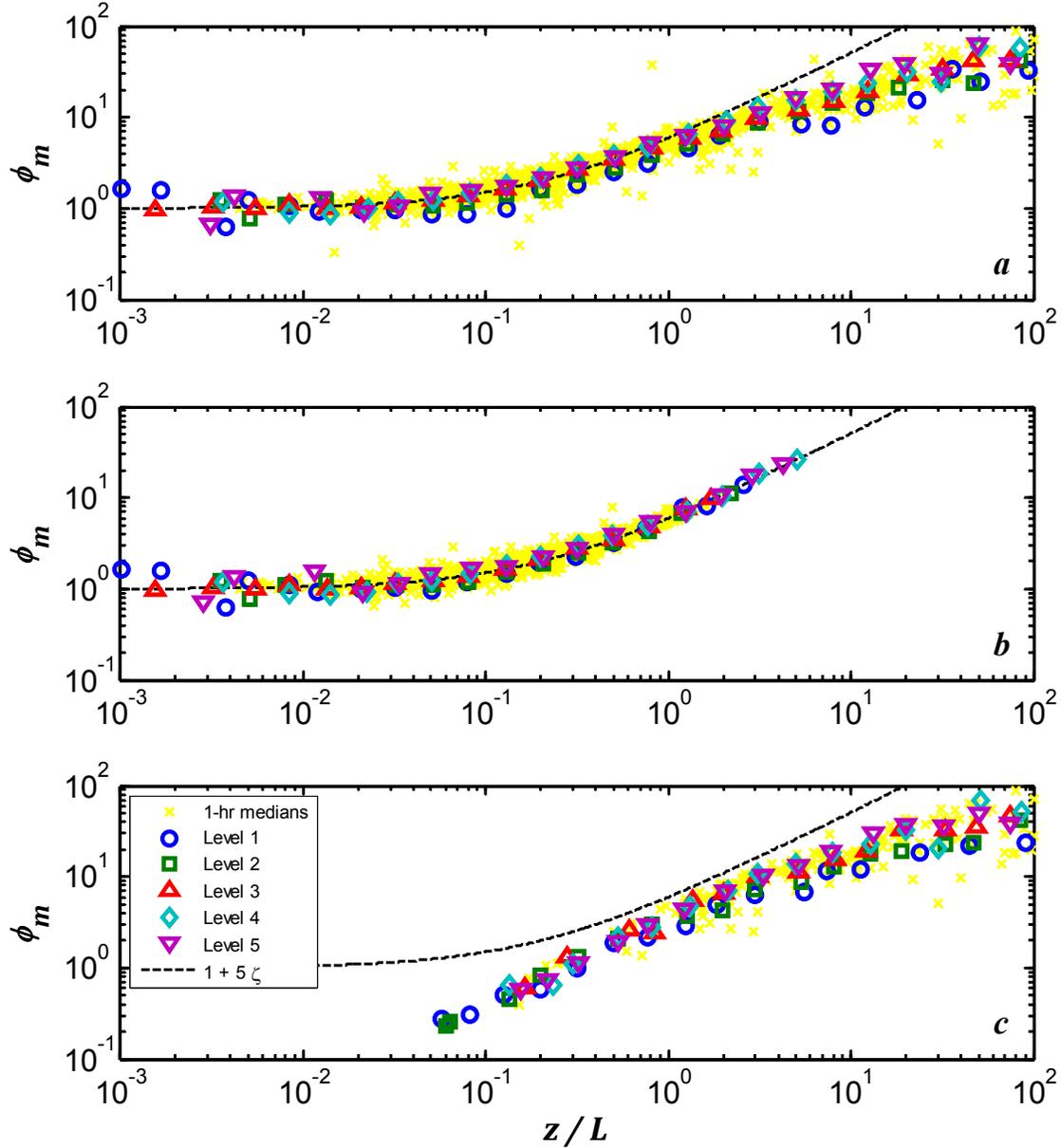

Fig. 10. The bin-averaged non-dimensional vertical gradient of mean wind speed $\varphi_m$ for five levels of the main SHEBA tower during the 11 months of measurements are plotted versus the Monin-Obukhov stability parameter for local scaling (*a*) for the original data, (*b*) in the subcritical regime when both prerequisites (10) with $Ri_{cr} = Rf_{cr} = 0.2$ have been imposed on the data, (*c*) in the supercritical regime when a prerequisite $Ri > Ri_{cr} = 0.2$ and $Rf > Rf_{cr} = 0.2$ has been imposed on the data. Individual 1-hr averaged SHEBA data based on the median fluxes for the five levels are shown as the background x-symbols.

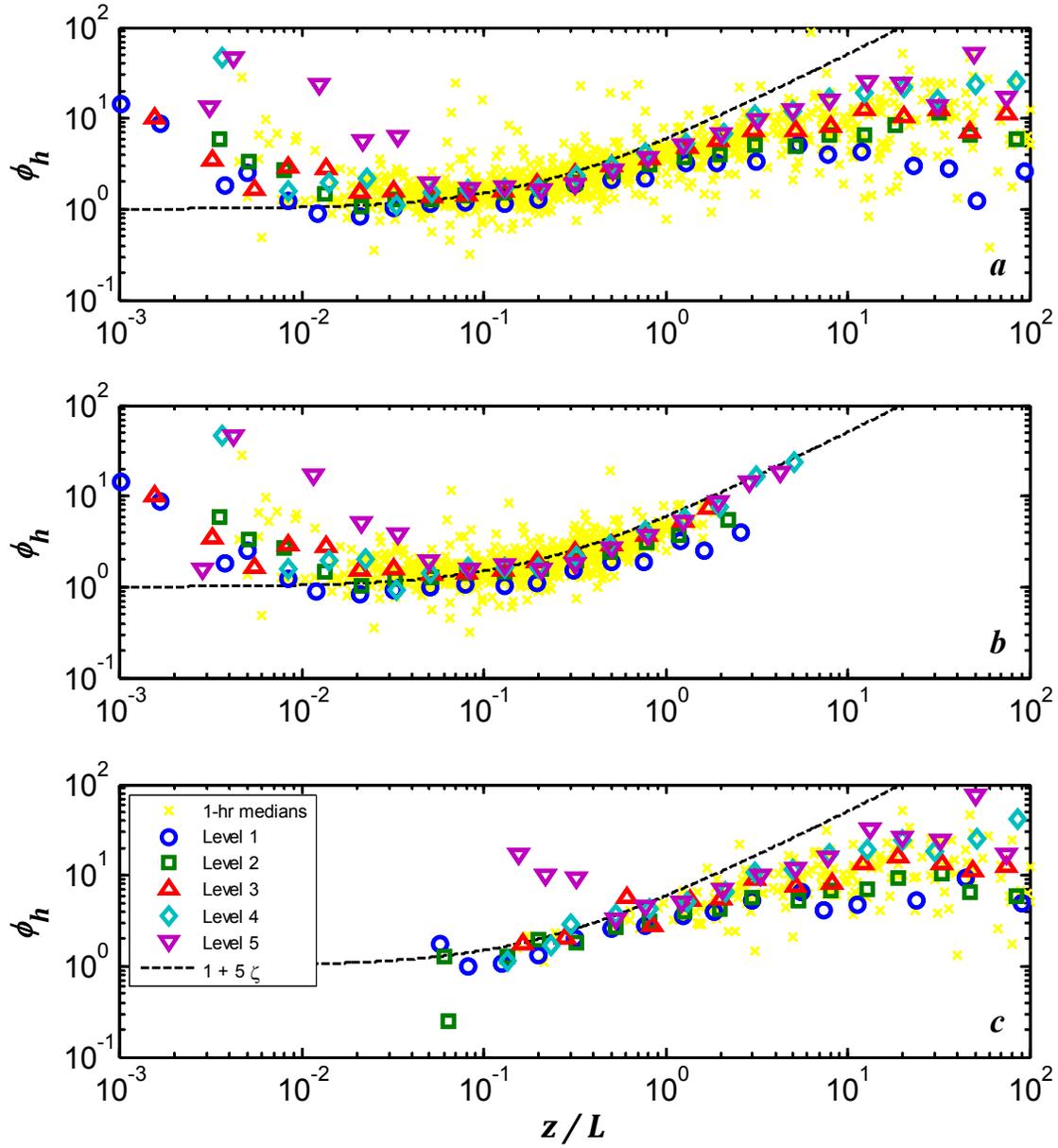

Fig. 11. Same as Fig. 10 but for the non-dimensional vertical gradient of mean potential temperature $\varphi_h$.

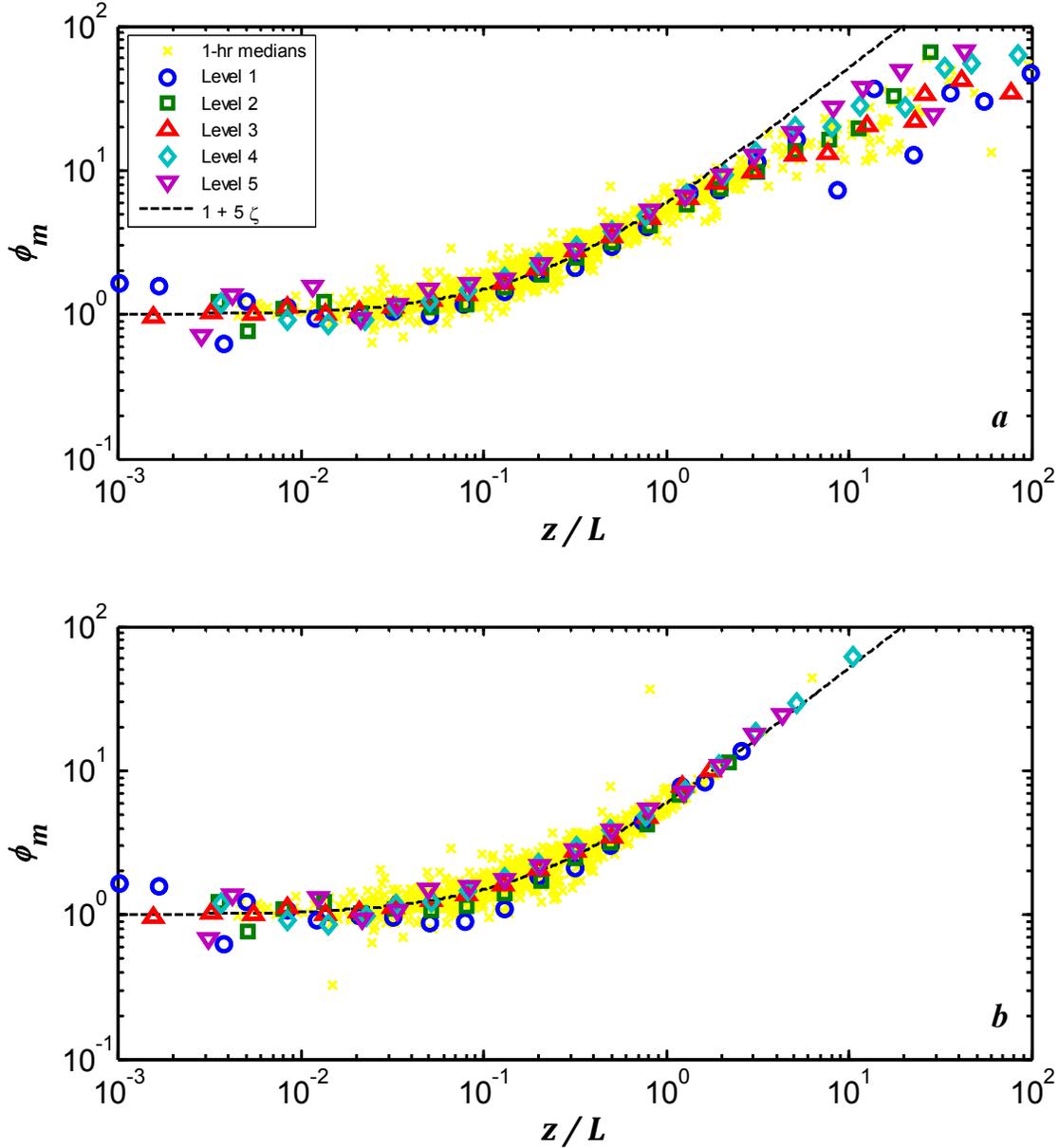

Fig. 12. The bin-averaged non-dimensional vertical gradient of mean wind speed $\varphi_m$ for five levels of the main SHEBA tower during the 11 months of measurements are plotted versus the local Monin-Obukhov stability parameter when (*a*) a single prerequisite (10a) with $Ri_{cr} = 0.2$ and (*b*) a single prerequisite (10b) with $Rf_{cr} = 0.2$ has been imposed on the data.

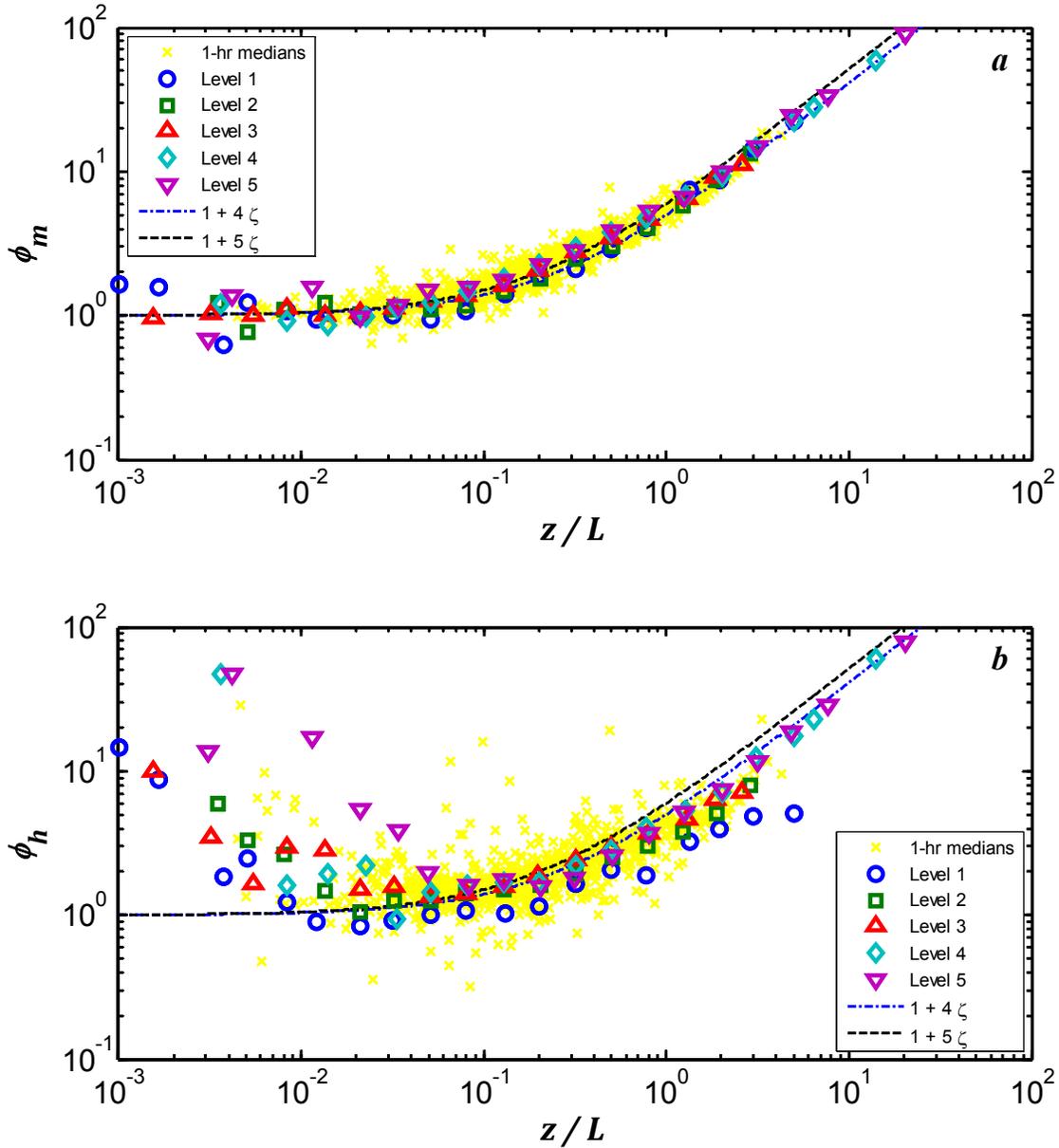

Fig. 13. The bin-averaged non-dimensional vertical gradients of (*a*) mean wind speed $\varphi_m$ and (*b*) mean potential temperature $\varphi_h$ for five levels of the main SHEBA tower during the 11 months of measurements are plotted versus the local Monin-Obukhov stability parameter in the subcritical regime when both prerequisites (10) with $Ri_{cr} = Rf_{cr} = 0.25$ has been imposed on the data.

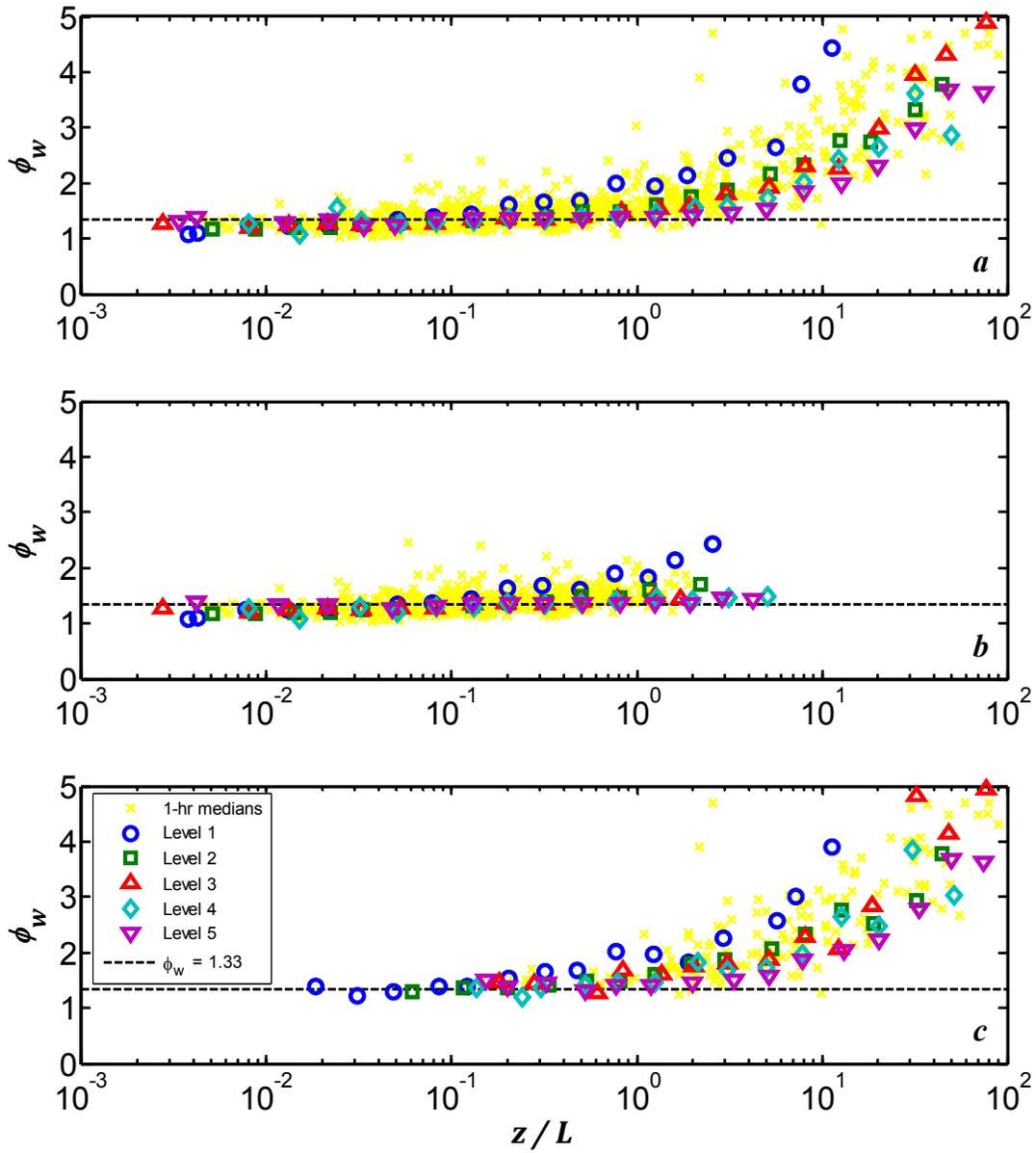

Fig. 14. Same as Fig. 10 but for the normalized standard deviation of the vertical wind speed component $\varphi_w = \sigma_w / u_*$ (local scaling). The horizontal dashed lines correspond to $\varphi_w = 1.33$.

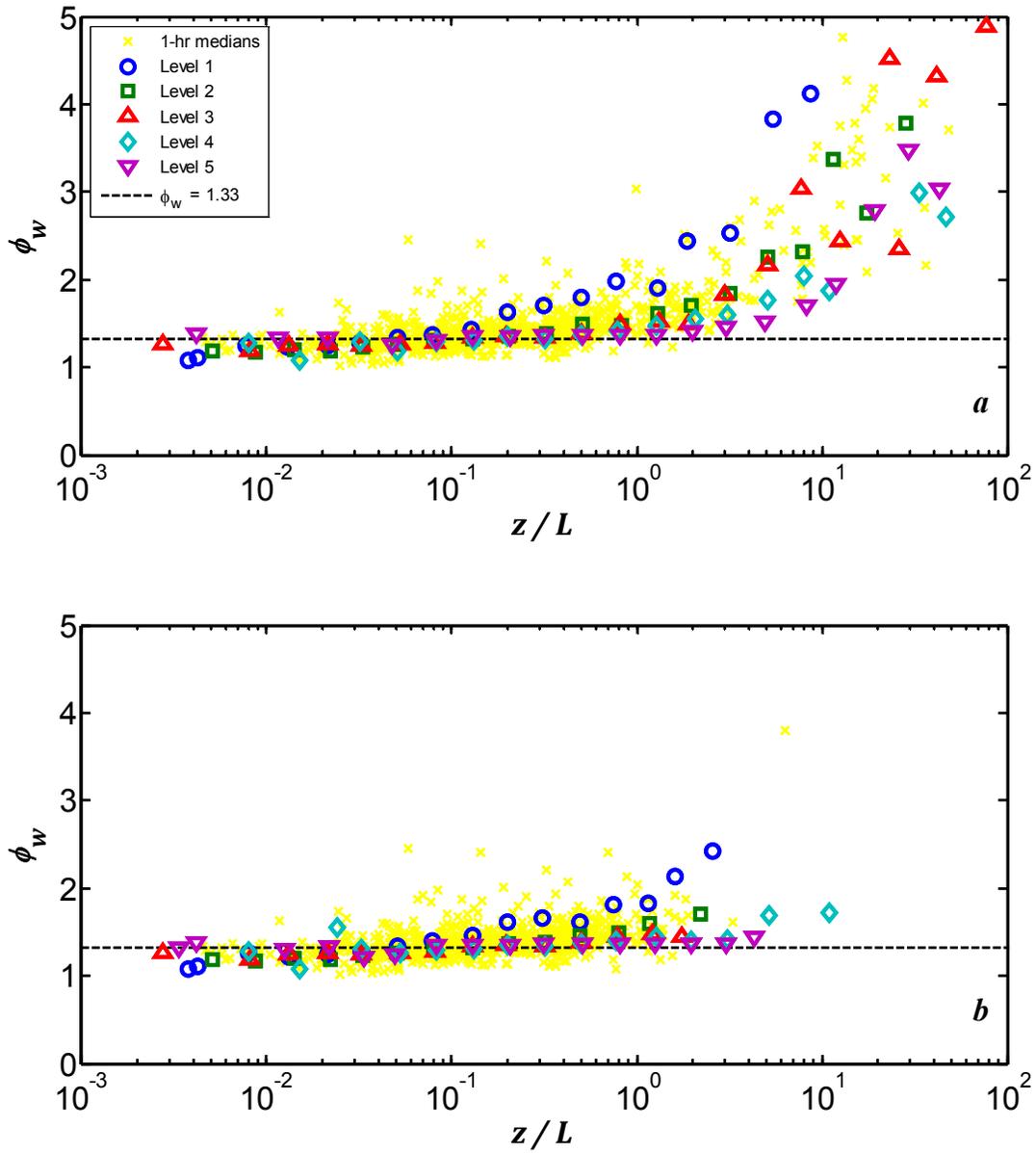

Fig. 15. Same as Fig. 12 but for the normalized standard deviation of the vertical wind speed component $\varphi_w = \sigma_w / u_*$. The horizontal dashed lines correspond to $\varphi_w = 1.33$.

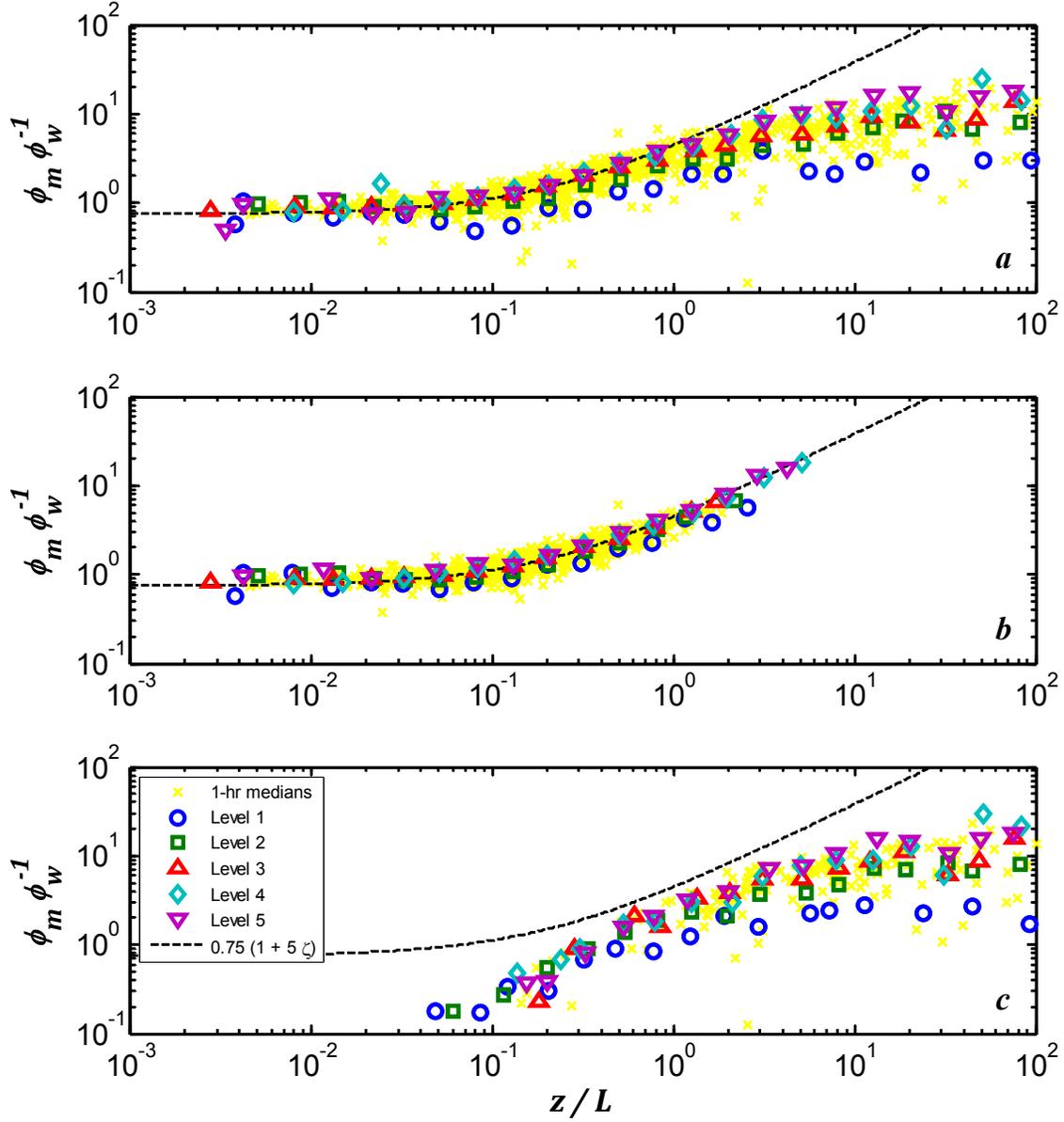

Fig. 16. Same as Fig. 10 but for the function $\varphi_m \varphi_w^{-1} = \left(\dfrac{\kappa z}{\sigma_w}\right)\dfrac{dU}{dz}$ which is a combination of the universal functions (3a) and (4a) for $\alpha = w$ and is not affected by the self-correlation. Symbols and notation are the same as in Fig. 10.